\documentclass[a4paper,11pt]{article}
\pdfoutput=1 

\usepackage{jheppub} 

\usepackage[T1]{fontenc} 

\usepackage{float}

\usepackage[dvipsnames]{xcolor}
\usepackage[utf8]{inputenc}
\usepackage{amsmath}
\usepackage{amsfonts}
\usepackage{slashed}
\usepackage{graphicx}
\usepackage{rotating}
\usepackage{enumitem}  
\usepackage{pifont}
\usepackage{comment}
\usepackage{appendix}
\usepackage{placeins}

\preprint{{\raggedleft IFIC/21-45 \par } {\raggedleft DO-TH 21/29 \par}}

\usepackage[caption=false]{subfig}

\newcommand{\Tr}{\mathrm{Tr}}

\newcommand{\imEtaU}{{\rm Im}(\eta_U)}

\newcommand{\imEtaQ}{{\rm Im}(\eta_Q)}
\newcommand{\reEtaU}{{\rm Re}(\eta_U)}

\newcommand{\reEtaQ}{{\rm Re}(\eta_Q)}

\usepackage{graphicx,amsmath,amssymb,epsfig,verbatim,mathrsfs,array,layout,textcomp,latexsym}
\usepackage{multirow}

\usepackage{tikz}
\usetikzlibrary{positioning,arrows}
\usetikzlibrary{decorations.pathmorphing}
\usetikzlibrary{decorations.markings}
\usepackage{pgfplots}
\pgfplotsset{compat=1.15}
\usepackage{xparse}
\usetikzlibrary{positioning,arrows,patterns}
\usetikzlibrary{decorations.markings}
\usetikzlibrary{calc}

\usepackage{ifthen}

\newcommand{\mpdiag}[2]{
	\hspace*{-0.5cm}\begin{minipage}[c]{0.36\textwidth}
		#1
	\end{minipage}
	\begin{minipage}[c]{0.61\textwidth}
		#2
	\end{minipage}
}

\newcommand{\mailto}[1]{\href{mailto:#1}{#1}}

\title{Electric dipole moments from colour-octet scalars}

\author[a,1]{Hector Gisbert,\note{\mailto{hector.gisbert@tu-dortmund.de}}}
\author[b,2]{Víctor Miralles,\note{\mailto{victor.miralles@ific.uv.es}}}
\author[b,3]{and Joan Ruiz-Vidal\note{\mailto{joan.ruiz@ific.uv.es}}}

\affiliation[a]{Fakultät für Physik, TU Dortmund, Otto-Hahn-Str.\,4, D-44221 Dortmund, Germany}
\affiliation[b]{IFIC, Universitat de València-CSIC, Apt. Correus 22085, E-46071 València, Spain }

\abstract{We present the contributions to electric dipole moments (EDMs) induced by the Yukawa couplings of an additional electroweak doublet of colour-octet scalars. The full set of one-loop diagrams and the enhanced higher-order effects from Barr-Zee diagrams are computed for the quark (chromo-)EDM, along with the two-loop contributions to the Weinberg operator. Using the stringent experimental upper limits on the neutron EDM, constraints on the parameter space of the Manohar-Wise model are derived.}

\begin{document} 

\maketitle

\flushbottom

\clearpage

\section{Introduction}\label{sec:intro}

The Standard Model (SM) has accurately anticipated a wide range of phenomena and satisfactorily described almost all experimental outcomes, resulting in the best description of nature that we have up to date. Nevertheless, it has clear shortcomings which require invoking new physics (NP). One of them is the observed baryon asymmetry of the universe, which differs by several orders of magnitude with respect to the SM prediction. To generate this asymmetry, the Sakharov conditions~\cite{Sakharov:1967dj} require additional sources of CP violation (CPV) beyond the SM that, in turn, can be constrained by the stringent experimental limits on the permanent electric dipole moment (EDM) of particles. Due to the small size of EDMs as predicted by the SM\footnote{For comprehensive reviews see for example Refs.~\cite{Pospelov:2005pr,Yamanaka:2013pfn}. }, any signal of a nonzero EDM in current or planned experiments would be an indisputable sign of NP. In this sense, these observables provide a background-free search for new sources of CPV beyond the SM. 

One of such sources can be found in extensions of the SM with additional scalar particles that transform as doublets under $SU(2)_L$ and as octets under $SU(3)_C$. These scalars were first proposed by Manohar and Wise (MW) \cite{Manohar:2006ga}, the original motivation being that they are one of the few scalar representations of the SM gauge group that can implement Minimal Flavour Violation (MFV) \cite{Chivukula:1987py,DAmbrosio:2002vsn}. In addition, these scalars emerge naturally with a mass of few TeVs from $SU(4)$, $SU(5)$ or $SO(10)$ unification theories at high energy scales~\cite{Georgi:1974sy,Georgi:1979df,Dorsner:2006dj,FileviezPerez:2013zmv,Perez:2016qbo,FileviezPerez:2019ssf,Bertolini:2013vta}. In this work, we study the phenomenology of EDMs arising from the Yukawa couplings of these scalars, complementing the extensive literature of phenomenological studies within this model~\cite{Gresham:2007ri, Gerbush:2007fe, Burgess:2009wm,Degrassi:2010ne, He:2011ti, Dobrescu:2011aa, Bai:2011aa, Arnold:2011ra, Kribs:2012kz, Reece:2012gi, Cao:2013wqa, He:2013tla, Cheng:2015lsa,Cheng:2016tlc, Martinez:2016fyd, Hayreter:2017wra,Cheng:2018mkc,Hayreter:2018ybt,Gisbert:2019ftm,Miralles:2019uzg,Miralles:2020tei,Eberhardt:2021ebh}.

\begin{figure}[h!]\label{fig:barr-zee-THDM}
    \centering
\vbox{
\resizebox{0.55\columnwidth}{!}{
\includegraphics[scale=1]{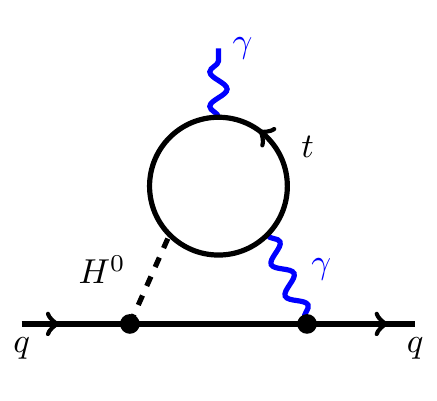}
\includegraphics[scale=1]{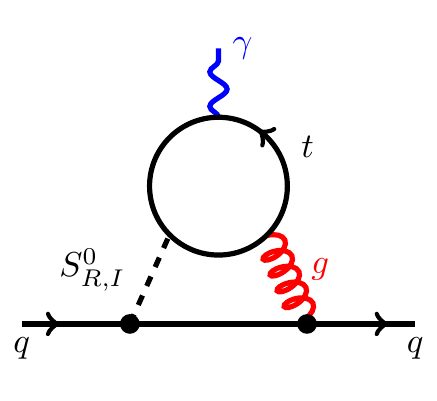}
}
}
\caption{Opposed to the THDM with colourless scalars (left), the leading contribution to the light quark EDM appears in the MW model through gluon exchange (right), enhancing the EDM by a factor $(C_F\,\alpha_s)/\alpha\sim 50$, with the couplings evaluated at the hadronic scale.}
\end{figure}

Only few publications have studied the effect of colour-octet scalars on the EDM of particles~\cite{Heo:2008sr,Fajfer:2014etr,Hisano:2012cc,Martinez:2016fyd}. The restrictions on their Yukawa couplings from heavy quark EDMs were studied in Ref.~\cite{Martinez:2016fyd} and numerically updated with more restrictive bounds in Ref.~\cite{Gisbert:2019ftm}. However, to derive robust limits on the model accounting for cancellation effects, the direct contributions to the neutron EDM through gluonic or light-quark operators need to be considered. The relevant contributions include two-loop diagrams which, to our knowledge, have not been computed in the literature within the MW model. 
%
%
Due to the colour structure of the new scalars, the light-quark EDMs are greatly enhanced when compared to the contribution from colourless scalars appearing in the two-Higgs-doublet model (THDM), as shown in Figure~\ref{fig:barr-zee-THDM}. This feature makes the hadronic EDMs powerful observables to assess the viability of the MW theory. 

This paper is organised as follows. First, in Section~\ref{sec:EFT}, we describe the relevant operators in the low-energy effective field theory that will contribute to the EDM observables. There, we also describe how to match the contributions of the new degrees of freedom to the operators at the hadronic scale. The contributions of these operators to the hadronic and nuclear EDMs, together with their current experimental limits, are presented in Section~\ref{sec:hadronEDMs}. In Section~\ref{sec:MWmodel} we briefly introduce the MW model. The main results of our paper are shown in Sections~\ref{sec:MWEDMcontrib} and \ref{sec:pheno}. First, in Section~\ref{sec:MWEDMcontrib}, we provide analytical expressions for the one- and two-loop contributions of coloured scalars to the chromo-EDM (CEDM) and EDM of the quarks, as well as to the Weinberg operator. Furthermore, we present a detailed comparison between the contributions to the quark EDM for each flavour. In Section~\ref{sec:pheno} we provide constraints on the CP-violating couplings of the colour-octet scalars from the neutron EDM, whose competitiveness with other observables constraining the same planes of the parameter space is demonstrated. Finally, in Section~\ref{sec:summary} we summarise our results.

\section{Effective theory framework}
\label{sec:EFT}

In this section, we provide the effective theory framework that describes the relevant operators contributing to the nuclear and atomic EDMs and their evolution to low energies through the renormalisation group.
Below the electroweak scale $\mu<\Lambda_{\text{EW}}$ the relevant flavour-diagonal CP-violating effective Lagrangian is given by\footnote{In the effective theory framework, we adopt the same conventions as in Ref.~\cite{Hisano:2012cc}.}
\begin{align}\label{eq:lagrangian}
\begin{split}
        \mathcal{L}_{\text{CPV}}\,=&\,\sum_{q}\,C_1^q(\mu)\,\mathcal{O}^q_1(\mu)\,+\,\sum_{q}\,C_2^q(\mu)\,\mathcal{O}^q_2(\mu)\,+\,C_3(\mu)\,\mathcal{O}_3(\mu)~,
\end{split}
\end{align}
where the sum of $q$ runs over all quark flavours. The effective operators are defined as
\begin{align}\label{eq:opEFF}
\begin{split}
    \mathcal{O}^q_1\,&=\,-\,\frac{i}{2}\,e\,\mathcal{Q}_q\,m_q\,(\bar{q}\,\sigma^{\mu\nu}\gamma_5\,q)\,F_{\mu\nu}~,\\
\mathcal{O}^q_2\,&=\,-\,\frac{i}{2}\,g_s\,m_q\,(\bar{q}\,\sigma^{\mu\nu}\,\gamma_5\,T^a\,q)\,G^a_{\mu\nu}~,\\
\mathcal{O}_3\,&=\,-\,\frac{1}{6}\,g_s\,f^{a b c}\,\epsilon^{\mu\nu\lambda\sigma}\,G^{a}_{\mu\rho}\,G^{b \rho}_{\nu}\,G^{c}_{\lambda\sigma}~.
\end{split}
\end{align}
Here, $F_{\mu\nu}$ and $G_{\mu\nu}^a$ with $a=1,...,8$ are the electromagnetic and gluon field strength tensors, $g_s$ is the strong coupling constant ($\alpha_s\equiv g_s^2/4\pi$), and $\sigma_{\mu\nu}\,=\,\frac{i}{2}[\gamma_\mu,\,\gamma_\nu]$. The matrix $T^a$ represents the generators of the $SU(3)_C$ group with normalisation $\Tr(T^a\,T^b)=\delta^{ab}/2$, and the tensor $f^{abc}$ the structure constant. The charge of up- and down-type quarks is $\mathcal{Q}_q=(2/3,-1/3)$. The expression for the covariant derivative will be relevant to define the anomalous dimension matrix later in Eq.~\eqref{eq:anomalousmatrix}. In this work, it is defined as $D_\mu\,=\,\partial_\mu\,-\,i\,e\,\mathcal{Q}_q\,A_\mu\,-\,i\,g_s\,G_\mu^a\,T^a$, where $A_\mu$ and $G_\mu^a$ are photon and gluon fields, respectively. We do not consider the contributions from four-quark operators since their moderate effect is well below the two-loop contributions as shown in Section~\ref{sec:fourquark}. The quark EDM $d_q(\mu)$, chromo-EDM $\widetilde{d}_q(\mu)$, and the usually defined coefficient $w(\mu)$ of the Weinberg operator are related to the Wilson coefficients by
\begin{align}
d_q(\mu)\:&=\:e\:\mathcal{Q}_q\:m_q(\mu)\:C_1^q(\mu) ~,\nonumber\\
\widetilde{d}_q(\mu)\:&=\:m_q(\mu)\:C_2^q(\mu)~,\\
w(\mu)\:&=\:-\:C_3(\mu)~.\nonumber
\end{align}
To compare the theory prediction with the low-energy EDM observables, these parameters need to be evaluated at the hadronic scale, $\mu_{\text{had}}\sim1\,$GeV. For that purpose, we employ the renormalisation group equations (RGEs),  
\begin{align}\label{eq:rge}
    \frac{\text{d}\,\overrightarrow{\mathcal{C}}(\mu)}{\text{d}\,\text{ln}\,\mu}\,=\,\widehat{\gamma}^{T}(\mu)\,\overrightarrow{\mathcal{C}}(\mu)\,,
\end{align}
where $\overrightarrow{\mathcal{C}}(\mu)=(C_1^q(\mu),\,C_2^q(\mu),\,C_3(\mu))$, and $\widehat{\gamma}(\mu)$ is the anomalous dimension matrix. At leading order in $\alpha_s$ it reads~\cite{Degrassi:2005zd,Dai:1989yh,Braaten:1990gq,Shifman:1976de,Brod:2018pli,Yamanaka:2017mef,Hisano:2012cc,deVries:2019nsu}
\begin{align} \label{eq:anomalousmatrix}
    \widehat{\gamma}(\mu)\,=\,\frac{\alpha_s(\mu)}{4\pi}\begin{pmatrix}
8\,C_F & 0 & 0 \\
8\,C_F & 16\,C_F\,-4\,N_C& 0\\
0 & 2\,N_C & N_C\,+\,2\,n_f\,+\,\beta_0
\end{pmatrix}\,,
\end{align}
where $C_F=(N_C^2-1)/(2 N_C)$, $\beta_0=(11 N_C -2 n_f)/3$ with $N_C=3$ and $n_f$ denotes the number of active flavours. Solving Eq.~\eqref{eq:rge}, we obtain the scale dependence of the Wilson coefficients $\overrightarrow{\mathcal{C}}(\mu)$ for a theory with constant number of active flavours. Starting at the NP scale $\mu=\Lambda_{\text{NP}}$, close to the top quark mass,\footnote{ The masses of the new scalars are in fact constrained to be above 1 TeV~\cite{Miralles:2019uzg,Eberhardt:2021ebh}. By simultaneously integrating out the new scalars and the top quark, we ignored the running in the range $[1\,\text{TeV},m_t]$. Conservatively, we estimate this effect to be at most of 30\%, which is smaller than the leading systematic error from the hadronic matrix elements in Eq.~\eqref{eq:edm_fun}.
} where the fundamental theory is matched to the effective one, the Wilson coefficients are evolved down to the bottom-quark mass scale with $n_f=5$. At this point, the bottom quark is integrated out, generating a threshold contribution of the bottom CEDM to the Weinberg operator as~\cite{Dai:1989yh,Boyd:1990bx,Dekens:2018bci}
\begin{align}\label{eq:thresholdweinberg}
    C_3(\mu_b^-)\,=\,C_3(\mu_b^+)\,+\,\frac{ \alpha_s(\mu_b^+)}{8\,\pi}\,C_2^q(\mu_b^+)~.
\end{align}
Here $\mu_b^+$ and $\mu_b^-$ refer to the scale $\mu_b\sim m_b$ in the theories with $n_f=5$ and $n_f=4$, respectively. Analogously, also the charm CEDM induces a threshold correction to the Weinberg operator, although it is numerically irrelevant for our study on the MW model. The final running with $n_f=4$ and $n_f=3$ brings the Wilson coefficients down to the hadronic scale $\mu_{\text{had}}\sim 1\,$GeV.

%

It is interesting to note that also the EDM of heavy quarks contributes to the hadronic EDMs~\cite{Gisbert:2019ftm}. This occurs through photon-loop corrections to the corresponding quark CEDM which, in turn, provides threshold corrections to the Weinberg operator, as mentioned above. 
If family-specific couplings exist that appear only in heavy quark EDMs, \textit{e.g.} in models of scalar Leptoquarks, this effect, suppressed in comparison to gluon-loop corrections, provides a unique window to access these family-specific couplings~\cite{Gisbert:2019ftm}. 
In the model considered in this work, we have checked that the leading constraints to the Yukawa couplings appear from the light quark (C)EDMs, as we will see in Section~\ref{sec:pheno}, and therefore we neglected $\mathcal{O}(\alpha)$ corrections in Eq.~\eqref{eq:anomalousmatrix} for the sake of simplicity.

\section{Matching onto nuclear and atomic EDMs}\label{sec:hadronEDMs}

The contributions of the operators in Eq.~\eqref{eq:opEFF} at the hadronic scale to the electric dipole moments of the neutron $d_n$, proton $d_p$, and mercury $d_{\text{Hg}}$ are computed with non-perturbative techniques of strong interactions at low energies. State-of-the-art coefficients for the CEDMs and Weinberg operator have been obtained in the literature with QCD sum rules~\cite{Pospelov:2000bw,Lebedev:2004va,Hisano:2012sc,Haisch:2019bml} and the quark model~\cite{Yamanaka:2020kjo}, while the contributions of the quark EDMs have been computed in lattice QCD~\cite{Bhattacharya:2015esa,Bhattacharya:2015wna,Bhattacharya:2016zcn,Gupta:2018qil,Gupta:2018lvp}, with significantly smaller errors. Assuming a Peccei-Quinn mechanism, these read~\cite{Dekens:2021bro}
\begin{align}\label{eq:edm_fun}
d_n&=g_T^u\,d_u+g_T^d\,d_d-(0.55\pm0.28)\,e\,\tilde d_u-(1.1\pm0.55)\,e\,\tilde d_d -20\,(1\pm0.5)\,{\rm MeV}\,e\,g_s\,w,\nonumber\\
d_{p}&=g_T^d\,d_u+g_T^u\,d_d+(1.30\pm0.65)\,e\, \tilde d_u+(0.60\pm0.30)\,e\,\tilde d_d +18\,(1\pm0.5)\,{\rm MeV}\,e\,g_s\,w\,,\nonumber\\
d_{\text{Hg}}&=-(2.1\pm 0.5)\cdot 10^{-4}\left[(1.9\pm0.1)d_n +(0.20\pm0.06) d_p \right]\,,
\end{align}
where the contributions to $d_{\rm Hg}$ from pion-nucleon interactions that are compatible with zero within $1\sigma$ have been left out. In these expressions, the tensor charges describing the light quark EDM contributions read $g_T^u = -0.213 \pm 0.012$ and $g_T^d = 0.820 \pm 0.029$. The current experimental limits of $d_n,d_p$, and $d_{\text{Hg}}$ are collected in Table~\ref{tab:bounds}.

\begin{table}[h!]
 \centering
 \begin{tabular}{lcc}
  \hline
  \hline
  Observable  &  Current bound & Future sensitivities \\
  \hline
   $d_n$ &  $1.8\,\cdot\,10^{-26}$  & $1.0\,\cdot\,10^{-28}$       \\
   $d_{p}$ & --   & $1.0\,\cdot\,10^{-29}$     \\
   $d_{\text{Hg}}$ & $6.3\,\cdot\,10^{-30}$   & $1.0\,\cdot\,10^{-30}$       \\
  \hline
  \hline
   \end{tabular}
   \caption{Current experimental limits (at 90\% C.L.) on the absolute value of the electric dipole moments of the neutron $d_n$~\cite{nEDM:2020crw}, and mercury $d_{\text{Hg}}$ \cite{Graner:2016ses}, and the future experimental sensitivity, including the proton $d_p$ in $e\,$cm units.
}
   \label{tab:bounds}
\end{table}

The current bound on the Mercury EDM~\cite{Graner:2016ses} is four orders of magnitude stronger than that of the neutron, which has been recently reported by the nEDM collaboration~\cite{nEDM:2020crw}. However, this difference is compensated by the suppression factor in front of $d_{n,p}$ in its contribution to $d_{\rm Hg}$, as shown in Eq.~\eqref{eq:edm_fun}.
As a result, we obtain almost identical constraints on the model parameters by using the $d_n$ or $d_{\rm Hg}$ experimental limits (within less than a 10\% difference), and in the numerical analysis of Section~\ref{sec:pheno} we will use only the direct limit on $d_n$.

\section{The scalar colour-octet model }\label{sec:MWmodel}

The inclusion of a scalar field transforming as $(8,\,2)_{1/2}$ under the SM gauge group $SU(3)_C \times SU(2)_L\times U(1)_Y$ allows the construction of more renormalisable invariant interactions~\cite{Manohar:2006ga}. In general, the Lagrangian describing colour-octet scalar interactions can be written as 
\begin{align}
    \mathcal{L}_{\text{MW}}\,=\,\mathcal{L}_{\text{SM}}\,+\,\mathcal{L}_{\text{kin}}\,+\,\mathcal{L}_{Y}\,+\,\mathcal{L}_{\text{S}}\,,
\end{align}
where $\mathcal{L}_{\text{SM}}$, $\mathcal{L}_{\text{kin}}$, $\mathcal{L}_{Y}$, and $\mathcal{L}_{\text{S}}$ represent the SM Lagrangian, the colour-octet scalar kinetic term, the interaction with SM fermions, and the scalar sector, respectively. 
The kinetic term
\begin{align}
 \mathcal{L}_{\text{kin}}=2\,\Tr[(D_\mu S)^\dagger D^\mu S]~,
 \label{eq:kin}
\end{align}
generates interactions with the SM gauge particles through the covariant derivative $D_\mu S=\partial_\mu S+ i\, g_s\,[G_\mu,S]\,+\,i\,g\, \widetilde{W}_\mu\, S+\frac{i}{2} \,g^\prime\,B_\mu \,S$. The scalar sector $\mathcal{L}_{\text{S}}$ encodes the interactions between the octet scalars which are given by~\cite{Manohar:2006ga} 
\begin{align}
\mathcal{L}_{\text{S}}&=2\,m_S^2\,\Tr(S^{\dagger i}S_i)+\lambda_1\,\phi^{\dagger i}\phi_i\,\Tr(S^{\dagger j}S_j)+\lambda_2\,\phi^{\dagger i}\phi_j\,\Tr(S^{\dagger j}S_i)\nonumber\\
&+\left[\lambda_3\,\phi^{\dagger i}\phi^{\dagger j}\,\Tr(S_i S_j)+\lambda_4\,\phi^{\dagger i}\, \Tr(S^{\dagger j} S_j S_i)+\lambda_5\,\phi^{\dagger i} \,\Tr(S^{\dagger j} S_i S_j)+\text{H.c.}\right]\label{eq:potential} \\&+\lambda_6\, \Tr(S^{\dagger i}S_iS^{\dagger j}S_j)+\lambda_7\, \Tr (S^{\dagger i}S_jS^{\dagger j}S_i)+\lambda_8\, \Tr(S^{\dagger i}S_i)\Tr( S^{\dagger j}S_j)\nonumber\\&+
\lambda_9 \,\Tr(S^{\dagger i}S_j)\Tr (S^{\dagger j}S_i)+\lambda_{10}\, \Tr( S_iS_j)\,\Tr( S^{\dagger i}S^{\dagger j})+\lambda_{11} \,\Tr(S_iS_j S^{\dagger j}S^{\dagger i})~,\nonumber
\end{align}
where $\phi = (\phi^+,\phi^0)^T$ is the usual Higgs doublet, and $\langle\phi^0\rangle=\frac{v}{\sqrt{2}}$ with $v\sim 246\,$GeV the vacuum expectation value (VEV). Here, $i$ and $j$ are $SU(2)_L$ indices, and all traces are in colour space. 
It can be seen from Eq.~\eqref{eq:potential} that $\lambda_3$, $\lambda_4$ and $\lambda_5$ can have complex phases. However, through an appropriate phase rotation of $S$, $\lambda_3$ can be chosen to be real. The phenomenological study of the present work will be limited to EDM observables arising from the CPV in the Yukawa couplings. The effect of the CP-violating parts of the potential will be considered in a future global-fit analysis using several observables at the same time.

The VEV, $v$, causes a mass splitting between the octet scalars. Decomposing the neutral octet scalars into two real scalars,
\begin{align}
    S^{a,0}\,=\,\frac{1}{\sqrt{2}}\,\left(S^{a,0}_R\,+\,i\,S^{a,0}_I\right)~,
\end{align}
one obtains the following relation between the physical masses
\begin{align}
m_{S^\pm}^2=m_S^2+\lambda_1\frac{v^2}{4}~,\hspace{1cm} m_{S^{0}_{R,I}}^2=m_S^2+(\lambda_1+\lambda_2\pm2\,\lambda_3)\,\frac{v^2}{4}~,
 \end{align}
where $m_{S^\pm}$ is the mass of the charged scalar, $m_{S^{0}_{R}}$ the mass of the neutral CP-even scalar and $m_{S^{0}_{I}}$ the mass of the CP-odd scalar.

Assuming MFV, the new Yukawa interaction can be parametrised by two complex numbers 
\begin{equation}\label{eq:yukawa_lagrangian}
\mathcal{L}_{Y}=-\sum^3_{i,j=1}\Big[\eta_D \,Y^d_{ij}\,\overline{Q}_{L_i}\,S\, d_{R_j}+\eta_U\, Y^u_{ij}\, \overline{Q}_{L_i}\, \widetilde{S}\,u_{R_j}+\text{h.c.}\Big]~,
\end{equation}
inducing new CP-violating sources beyond the SM that contribute to the (C)EDM of quarks. In Eq.~\eqref{eq:yukawa_lagrangian}, $Q_{L}$ represents the left-handed quark doublet, and $u_{R}$ and $d_R$ correspond to the right-handed up- and down-quarks singlets, respectively. The new scalar fields are written as $S=S^a\,T^a$ with $S^a = (S^{a,+},\, S^{a,0})^T$. In Eq.~\eqref{eq:yukawa_lagrangian}, the shorthand notation $\widetilde{S}=i\, \sigma_2\,S$ is employed, where $\sigma_2$ is the usual Pauli matrix.

\section{Contributions to EDMs in the colour-octet scalar model}\label{sec:MWEDMcontrib}

In this section, we analyse the different contributions to the neutron EDM from the colour-octet scalars. Namely, we will derive the expressions for the quark (C)EDM at one-loop level and the enhanced contributions at two-loop level, together with the leading two-loop contributions to the Weinberg operator. The constraints on the model using these expressions are discussed in Section~\ref{sec:pheno}.

\subsection{One-Loop contributions}\label{sec:oneloop}

\begin{figure}[h!]
    \centering
 
\vbox{
\resizebox{1.0\columnwidth}{!}{
\subfloat[][]{\includegraphics[scale=1]{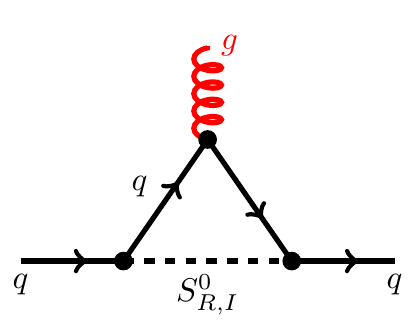}}
\subfloat[][]{\includegraphics[scale=1]{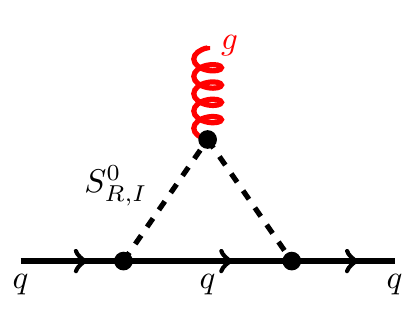}}
\subfloat[][]{\includegraphics[scale=1]{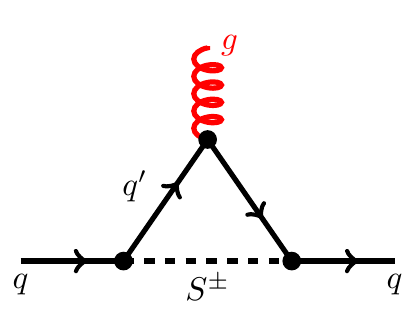}}
\subfloat[][]{\includegraphics[scale=1]{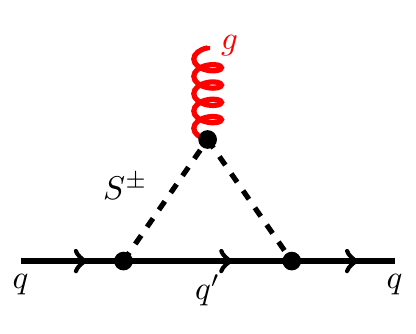}}
}
}
    \caption{Neutral $S^0$ and charged $S^\pm$ scalars contributing to the CEDM of a quark $q$.}
    \label{fig:oneloopCEDM}
\end{figure}

\begin{figure}[h!]
\centering

\vbox{
\resizebox{0.75\columnwidth}{!}{
\subfloat[][]{\includegraphics[scale=1]{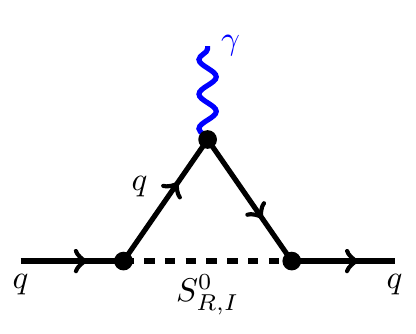}}
\subfloat[][]{\includegraphics[scale=1]{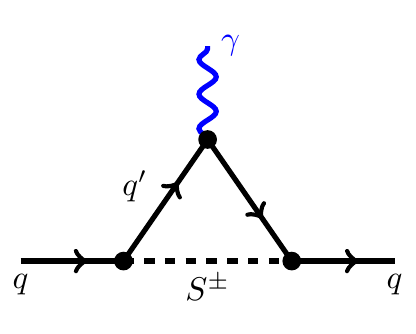}}
\subfloat[][]{\includegraphics[scale=1]{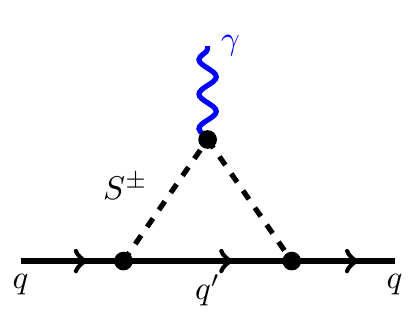}}
}
}
\caption{Neutral $S^0$ and charged $S^\pm$ scalars contributing to the EDM of a quark $q$. }\label{fig:oneloopEDM}
\end{figure}

At one-loop level, the (C)EDM of a quark $q$ receives contributions from neutral and charged scalars, $S^0_{I,R}$ and $S^\pm$, as shown in Figure~\ref{fig:oneloopEDM} (Figure~\ref{fig:oneloopCEDM}). These contributions can be computed using standard techniques, and are finite since the Lagrangian does not contain any tree-level (C)EDM. Our results for the CEDM of a quark $q$ reads
\begin{align}
\begin{split}
    \widetilde{d}_q\,=\,\text{sgn}(\mathcal{Q}_q)\left(\mathcal{N}_{(a)}^{qg}\,\widetilde{d}^{\,\text{(a)}}_q+\mathcal{N}_{(b)}^{qg}\,\widetilde{d}^{\,\text{(b)}}_q+\mathcal{N}_{(c)}^{qg}\,\widetilde{d}^{\,\text{(c)}}_q+\mathcal{N}_{(d)}^{qg}\,\widetilde{d}^{\,\text{(d)}}_q\right)\,,
\end{split}
\end{align}
with\footnote{Eqs.~\eqref{eq:CEDM_oneloop} and \eqref{eq:EDM_oneloop} can be translated to the basis of Ref.~\cite{Martinez:2016fyd} through the following replacements $d_q\to-\,d_q$ and $\widetilde{d}_q\to-\,\widetilde{d}_q$, as well as $\eta_U\to\eta_U\,\text{e}^{i(\alpha_U+\pi)}$ and $\eta_D\to\eta_D\,\text{e}^{-i\alpha_D}$.} 
\begin{align}\label{eq:CEDM_oneloop}
\begin{split}
    \widetilde{d}^{\,\text{(a)}}_{q}\,&=\,-\, \frac{G_F}{\sqrt{2}}\,\frac{m_{q}^3}{4\,\pi^2}\,\text{Re}(\eta_Q)\,\text{Im}(\eta_Q)\,\left(\frac{F_{2,0}(r_{q R})}{m_{S_R^0}^2}\,-\,\frac{F_{2,0}(r_{q I})}{m_{S_I^0}^2}\right)\,,\\
    \widetilde{d}^{\,\text{(b)}}_{q}\,&=\,-\,
    \frac{G_F}{\sqrt{2}}\,\frac{m_{q}^3}{4\,\pi^2}\,\text{Re}(\eta_Q)\,\text{Im}(\eta_Q)\,\left(\frac{F_{1,1}(r_{q R})}{m_{S_R^0}^2}\,-\,\frac{F_{1,1}(r_{q I})}{m_{S_I^0}^2}\right)\,,\\
    \widetilde{d}^{\,\text{(c)}}_{q}\,&=\,\frac{G_F}{\sqrt{2}}\,\frac{m_{q}\,m_{q^\prime}^2\,|V_{q q^\prime}|^2}{4\,\pi^2}\,\bigg(\text{Re}(\eta_Q)\, \text{Im}(\eta_{Q^\prime})-\text{Re}(\eta_{Q^\prime}) \,\text{Im}(\eta_Q)\bigg)\,\left(\frac{G_{2,0}(r_{q},r_{q^\prime})}{m_{S^\pm}^2}\right)\,,\\
    \widetilde{d}^{\,\text{(d)}}_{q}\,&=\,\frac{G_F}{\sqrt{2}}\,\frac{m_{q}\,m_{q^\prime}^2\,|V_{q q^\prime}|^2}{4\,\pi^2}\,\bigg(\text{Re}(\eta_Q)\, \text{Im}(\eta_{Q^\prime})-\text{Re}(\eta_{Q^\prime}) \,\text{Im}(\eta_Q) \bigg)\,\left(\frac{G_{1,1}(r_{q},r_{q^\prime})}{m_{S^\pm}^2}\right)\,,
\end{split}
\end{align}
where
\begin{align}\label{eq:colorfactors}
    \mathcal{N}_{(a)}^{qg}\,=\,\mathcal{N}_{(c)}^{qg}\,=\,-\frac{C_A - 2 \,C_F}{2}\,=\,-\,\frac{1}{6}\,,\quad
    \mathcal{N}_{(b)}^{qg}\,=\,\mathcal{N}_{(d)}^{qg}\,=\,\frac{C_A}{2}\,=\,\frac{3}{2}\,,
\end{align}
are colour factors emerging from the color structures appearing in the Feynman diagrams $T^{b} T^{a} T^{b}=-\frac{(C_A - 2C_F)}{2}\,T^{a}$ and $f^{a b c} \,T^{b} T^{c} = i\,\frac{C_A}{2}\,T^{a}$ with $C_A=3$.\footnote{Sum over repeated (colour) indices is understood here and in the following. } The loop functions $F_{n,m}(r)$ and $G_{n,m}(r_1,r_2)$ are defined in the notation of Ref.~\cite{Martinez:2016fyd} as
\begin{align}
    F_{n,m}(r)\,&=\,\int_{0}^1\,\frac{x^n\,(1-x)^m}{1\,-\,x\,+\,r^2\,x^2}\,\text{d}x\,,\\
    G_{n,m}(r_1,r_2)\,&=\,\int_{0}^1\,\frac{x^n\,(1-x)^m}{(1-x)(1\,-\,r_1^2\,x)\,+\,r_2^2\,x^2}\,\text{d}x\,,
\end{align}
with $r_q\equiv\frac{m_q}{m_{S^\pm}}$, $r_{qR}\equiv\frac{m_q}{m_{S^0_R}}$, and $r_{qI}\equiv\frac{m_q}{m_{S^0_I}}$.

While both neutral and charged scalars couple to gluons, only charged scalars couple to photons, making the CEDM receive an extra contribution with respect to the EDM, depicted in diagram (b) of Figure~\ref{fig:oneloopCEDM}. The contributions to the EDM of a quark $q$, shown in Figure~\ref{fig:oneloopEDM}, share the same loop functions with the CEDM diagrams. Therefore, the results for the EDM can be given in terms of the CEDM expressions (Eqs.~\eqref{eq:CEDM_oneloop}) as

\begin{align}
\begin{split}
    d_{q}\,=\,\text{sgn}(\mathcal{Q}_q)\left(\mathcal{N}_{(a)}^{q\gamma}\,d^{\,\text{(a)}}_{q}+\mathcal{N}_{(b)}^{q\gamma}\,d^{\,\text{(b)}}_{q}+\mathcal{N}_{(c)}^{q\gamma}\,d^{\,\text{(c)}}_{q}\right)\,,
\end{split}
\end{align}
where 
\begin{align}\label{eq:EDM_oneloop}
    d^{\,\text{(a)}}_{q}\,=\,e\,\mathcal{Q}_q\,\widetilde{d}^{\,\text{(a)}}_{q}\,,\quad
    d^{\,\text{(b)}}_{q}\,=\,e\,\mathcal{Q}_{q^\prime}\,\widetilde{d}^{\,\text{(c)}}_{q}\,,\quad
    d^{\,\text{(c)}}_{q}\,=\,e\,(\mathcal{Q}_q\,-\,\mathcal{Q}_{q^\prime})\,\widetilde{d}^{\,\text{(d)}}_{q}\,,
\end{align}
and
\begin{align}
    \mathcal{N}_{(a)}^{q\gamma}\,=\,\mathcal{N}_{(b)}^{q\gamma}\,=\,\mathcal{N}_{(c)}^{q\gamma}\,=\,C_F\, .
\end{align}
 The colour factor $C_F$ emerges from the combination of two colour matrices $ (T^a T^a)_{ij}=C_F\,\delta_{ij}$, each of them provided by one of the two Yukawa couplings in the diagrams of Figure~\ref{fig:oneloopEDM}.

Correcting by colour factors, our results are in good agreement with the literature of the colourless THDM~\cite{Iltan:2001vg,Jung:2013hka}. Additionally, for the diagrams that do not appear in the THDM but emerge through the introduction of colour-octet scalars (see Figure~\ref{fig:oneloopCEDM} (b) and (d)) we found agreement with the previous calculation in Ref.~\cite{Martinez:2016fyd}. However, in this reference, the loop function $F_{2,0}$ in Eqs.~\eqref{eq:CEDM_oneloop} is replaced by $F_{0,0}$, which differs from our results and those of Refs.~\cite{Iltan:2001vg,Jung:2013hka}.

\subsection{Two-loop contributions}\label{sec:twoloop}

In the previous section, we have studied all one-loop contributions to the (C)EDM of the quarks. Since light quark (C)EDMs are heavily suppressed at one-loop level by powers of the quark masses, the leading contributions to the neutron EDM will happen at two-loop level. In the following, we derive these two-loop contributions to the quark (C)EDM and the Weinberg operator.

\subsubsection{Barr-Zee diagrams}

\begin{figure}[h]
    \centering
\vbox{
\resizebox{0.95\columnwidth}{!}{
\subfloat[][]{\includegraphics[scale=1]{diagrams/barzeeS0gluontphoton}}
\subfloat[][]{\includegraphics[scale=1]{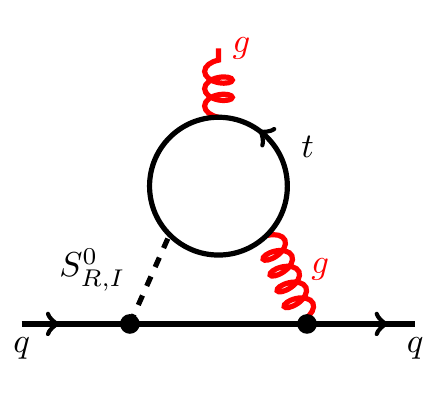}}
\subfloat[][]{\includegraphics[scale=1]{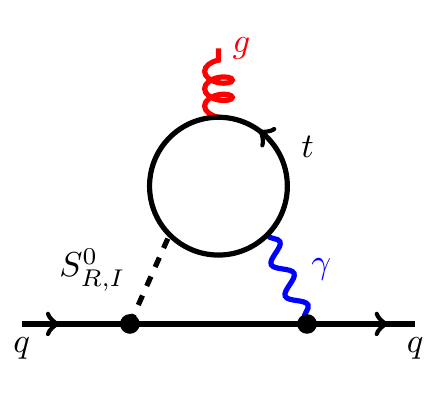}}
\subfloat[][]{\includegraphics[scale=1]{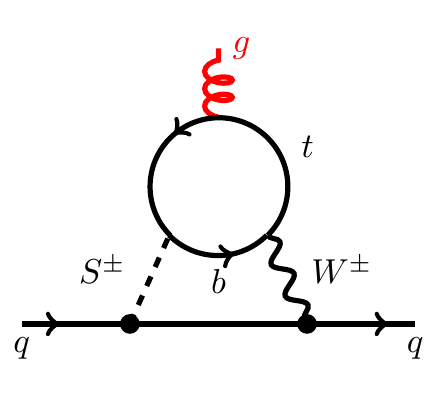}}
}
}
\caption{Barr-Zee type diagrams contributing to the CEDM (diagram (a)) and to the EDM (diagrams (b)) in the MW model. 
}\label{fig:barr_zee}
\end{figure}

Although being suppressed by additional coupling constants and loop factors, the Barr-Zee type diagrams, shown in Figure~\ref{fig:barr_zee}, benefit from the enhancement of the top-quark Yukawa coupling in flavour models. 
Among the contributions to the CEDM, the one depicted by diagram (b) (Figure~\ref{fig:barr_zee}) largely dominates, since it is enhanced by the strong coupling constant (from the internal gluon propagator) and it is not suppressed by any mass of the electroweak bosons. For this reason, this will be the only diagram that will be included in the expressions of the CEDM. 
Contributing to the quark EDM, only the diagram (a) (Figure~\ref{fig:barr_zee}) is numerically relevant.

The Barr-Zee contribution to the CEDM is given by
\begin{align}
\begin{split}
    \widetilde{d}_{q}^{\text{BZ}} =\,-\, 2\,\sqrt{2}\,G_F\frac{\alpha_s \,m_q }{(4\pi)^3}\, \mathcal{N}^{qg}_{\text{BZ}}\,
    \bigg( 
    &\imEtaQ \reEtaU \mathcal{F}^{(1)}\left(r_{tR}\right)
    +\reEtaQ\imEtaU\mathcal{F}^{(1)}\left(r_{tI}\right)   \\
    + &\imEtaU \reEtaQ \widetilde{\mathcal{F}}^{(1)}\left(r_{tR}\right)
    +\imEtaQ\reEtaU \widetilde{\mathcal{F}}^{(1)}\left(r_{tI}  \right)\bigg),
\end{split}  
\end{align}
where we have defined the loop functions\footnote{$\mathcal{F}^{(1)}$ and $\widetilde{\mathcal{F}}^{(1)}$ are related to $f$ and $g$ from Ref.~\cite{Jung:2013hka}, through $\mathcal{F}^{(1)}(\sqrt{r})=-f(r)$ and $\widetilde{\mathcal{F}}^{(1)}(\sqrt{r})=g(r)$~.
}
\begin{align}
&\mathcal{F}^{(1)}(r)=\frac{r^2}{2}\int_0^1 \text{d}x\, \frac{2x(1-x) -1}{r^2 - x(1-x)} \log \frac{r^2}{x(1-x)}\,,\\
    &\widetilde{\mathcal{F}}^{(1)}(r)=\frac{r^2}{2}\int_0^1 \text{d}x \, \frac{1}{r^2 - x(1-x)} \log \frac{r^2}{x(1-x)}\,,
\end{align}
and the colour factor $\mathcal{N}^{qg}_{\text{BZ}}=\frac{N_C^3-N_C-4}{4N_C}$ comes from $\Big(\Tr(T^a T^b T^c)+\Tr(T^b T^a T^c)\Big)T^b T^c=\frac{d^{a b c}}{2} T^b T^c=\frac{N_C^3-N_C-4}{4N_C} T^a$. The first two traces correspond to the current flow of the top quark in the loop, clockwise and counterclockwise. 
Similar to the one-loop case, the EDM depends on the same loop functions as the CEDM and they are related by 
\begin{align}
     d_{q}^{\text{BZ}} = e\, \frac{\mathcal{Q}_t\,C_F}{\mathcal{N}^{qg}_{\text{BZ}}} \,
    \widetilde{d}_{q}^{\text{BZ}}\,,
\end{align}
with $\mathcal{Q}_t=\frac{2}{3}$ being the charge of the top quark. For details on the computation of these diagrams see Appendix~\ref{app:BarrZee}.

\subsubsection{Weinberg contribution}

 \begin{figure}[h!] 
    \centering
\vbox{
\resizebox{0.70\columnwidth}{!}{
\subfloat[][]{\includegraphics[scale=1]{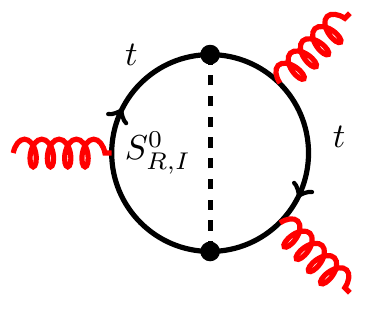}}
\subfloat[][]{\includegraphics[scale=1]{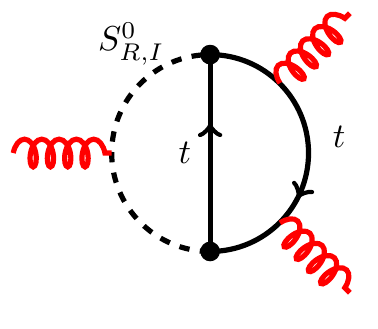}}
\subfloat[][]{\includegraphics[scale=1]{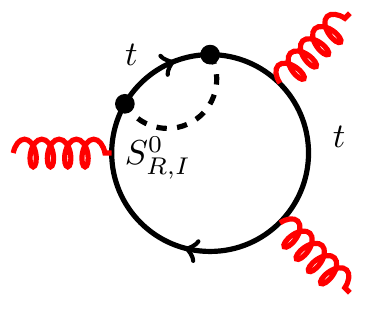}}
}
}
   \caption{Neutral scalar contributions to the Weinberg operator. The colour structure of diagram (a) yields a suppression factor of $1/6$ with respect to a THDM with colour-singlet scalars. In turn, diagram (b) is specific to colour-octet scalars, and diagram (c) vanishes. 
   }
    \label{fig:twoloopWeinbergNeut}
\end{figure}

 \begin{figure}[h!] 
    \centering
\vbox{
\resizebox{0.9\columnwidth}{!}{

\subfloat[][]{\includegraphics[scale=1]{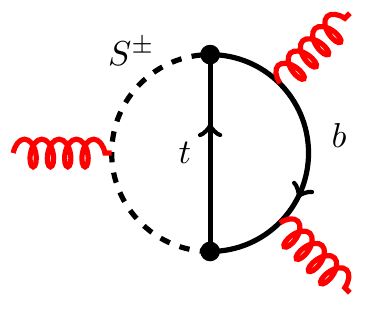}}
\subfloat[][]{\includegraphics[scale=1]{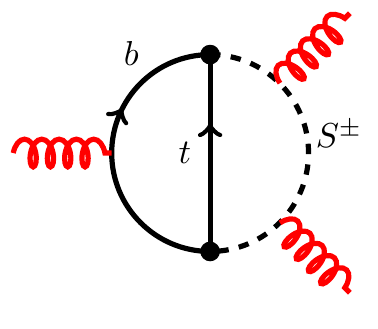}}
\subfloat[][]{\includegraphics[scale=1]{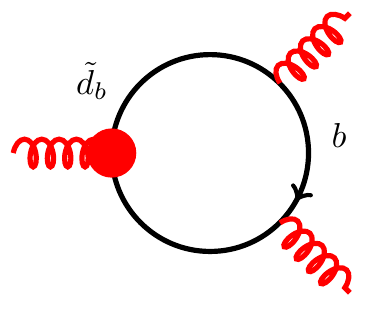}}
\subfloat[][]{\includegraphics[scale=1]{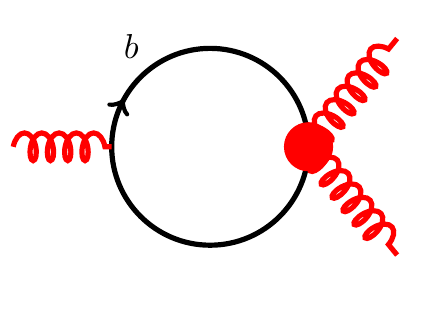}}

}
}
   \caption{Charged scalar contributions to the Weinberg operator. Below the top quark mass scale, diagrams (a) and (b) are accounted for through the effective operators of the bottom quark, depicted as red circles in diagrams (c) and (d). These induce a threshold correction to the Weinberg operator at the bottom quark mass scale, as shown in Eq.~\ref{eq:thresholdweinberg}.
   }
    \label{fig:twoloopWeinbergCharg}
\end{figure}

The Weinberg operator can also play an important role since it contributes directly to the neutron EDM and does not suffer from light quark mass suppressions. Furthermore, it also has an impact on the light quarks' (C)EDM due to the operator mixing in the RGEs. The first-order contribution appears at the two-loop level via the exchange of neutral (Figure~\ref{fig:twoloopWeinbergNeut}) or charged (Figure~\ref{fig:twoloopWeinbergCharg}) scalars. For the diagrams with neutral scalars, the masses of the top quark and new scalars running in the loop are assumed to be of the same order and, therefore, the complete two-loop diagrams must be calculated, yielding\footnote{Notice that Wilson coefficient of the Weinberg operator from Ref.~\cite{Jung:2013hka} can be translated to our basis through $C_W\,=\,-\,g_s\,C_3$.}
\begin{align}\label{eq:loopWeinberg_neutral}
\begin{split}
    w^{\,\text{(a)}}\,&=\,4\,\sqrt{2}\,G_F\,\frac{\alpha_s}{(4\pi)^3}
    \,\imEtaU\,\reEtaU\,\mathcal{N}_{(a)}^w  \, \big( h(r_{tR}) - h(r_{tI}) \big)\,,\\
    w^{\,\text{(b)}}\,&=\,4\,\sqrt{2}\,G_F\,\frac{\alpha_s}{(4\pi)^3}\,\imEtaU\,\reEtaU\,\mathcal{N}_{(b)}^w  \, \big( g(r_{tR}) - g(r_{tI}) \big)\, .
\end{split}
\end{align}
Here $\mathcal{N}_{(a)}^w=\mathcal{N}_{(b)}^w=-\frac{C_A - 2 \,C_F}{2}$ are the colour factors emerging from $ T^{b} T^{a} T^{b}$, similar to Eq.~\eqref{eq:colorfactors}, and
\begin{align}
\begin{split}
    h(r)&=\frac{r^4}{4}\int_0^1 \text{d}x \int_0^1 \text{d}y \frac{y^3x^3(1-x)}{[r^2 x (1-y x)+(1-y)(1-x)]^2}\,,\\
    g(r)&=\frac{r^4}{4}\int_0^1 \text{d}x \int_0^1 \text{d}y \frac{y^3x^2(1-x)^2}{[r^2 x (1-y x)+(1-y)(1-x)]^2}\,,  
\end{split}
\end{align}
are the corresponding loop functions. In turn, the diagram (c) of Figure~\ref{fig:twoloopWeinbergNeut} vanishes. The details of the calculation are found in Appendix~\ref{app:wein}. Notice that $h(r)$ corresponds to the well-known Weinberg loop function of Ref.~\cite{Weinberg:1989dx} but $g(r)$ is only appearing for coloured scalars since they can couple to gluons. Looking at Eq.~\eqref{eq:loopWeinberg_neutral} we see how, as it happened for the one-loop contribution to the (C)EDMs of the quarks, there is a relative minus sign between the contribution of the CP-even and CP-odd neutral scalars. Therefore, this contribution will be suppressed by the mass splitting of the neutral scalars which, as can be seen in Ref.~\cite{Eberhardt:2021ebh}, is around two orders of magnitude smaller than the mass of the scalars. This suppression is large and, for the numerical analysis in Section~\ref{sec:pheno}, we will assume that all scalar masses are the same, effectively neglecting the contribution of the neutral scalars in the Weinberg operator.

The calculation of the charged scalar contributions, shown in Figure~\ref{fig:twoloopWeinbergCharg}, proceeds differently. Having a bottom quark propagator in the loop, these diagrams will only induce a contribution to the Weinberg operator below the bottom-quark mass scale. At this scale, the NP particles and the top quark have already been integrated out and their information is encoded in the effective vertices shown in diagrams (c) and (d) of Figure~\ref{fig:twoloopWeinbergCharg}. These one-loop diagrams generate a threshold contribution to the Weinberg operator from the bottom CEDM $\tilde d_b$, as shown in Eq.~\eqref{eq:thresholdweinberg}. The main contribution to $\tilde d_b$ comes from the one-loop diagrams 
(c) and (d) of Figure~\ref{fig:oneloopCEDM}, which are not suppressed by any kind of light quark mass or CKM factor.

\subsection{Four-quark contributions}\label{sec:fourquark}

\begin{figure}[h!]
\centering
\vbox{
\includegraphics[scale=1]{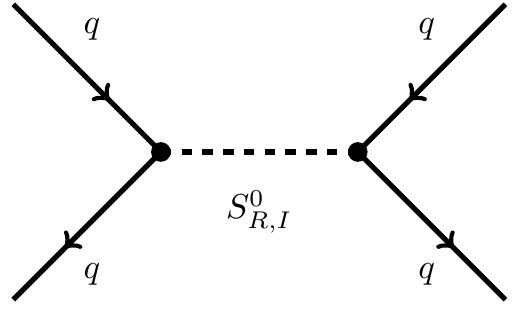}
}
\caption{Four-quark contribution from the neutral scalars $S^0_{R,I}$.}\label{fig:4quark}
\end{figure}

Four-quark interactions due to the exchange of colour-octet scalars (Figure~\ref{fig:4quark}) have been studied in detail in Ref.~\cite{Hisano:2012cc}. Using the results from that work, we observe that they represent a sub-leading contribution to the neutron EDM 
\begin{align}
    d_n^{(4\,q)}\,\sim\,3\,\cdot\,10^{-29}\,e\,\text{cm}\,\imEtaQ\, \reEtaQ\,\left(\frac{1\,\text{TeV}}{m_{S_{R,I}^0}}\right)^2~,
\end{align}
well below the two-loop contributions presented in Section~\ref{sec:twoloop}. Therefore, we neglect them in the following.

\subsection{Discussion }

\begin{figure}[h!]
    \centering
    \includegraphics[width=0.46\textwidth]{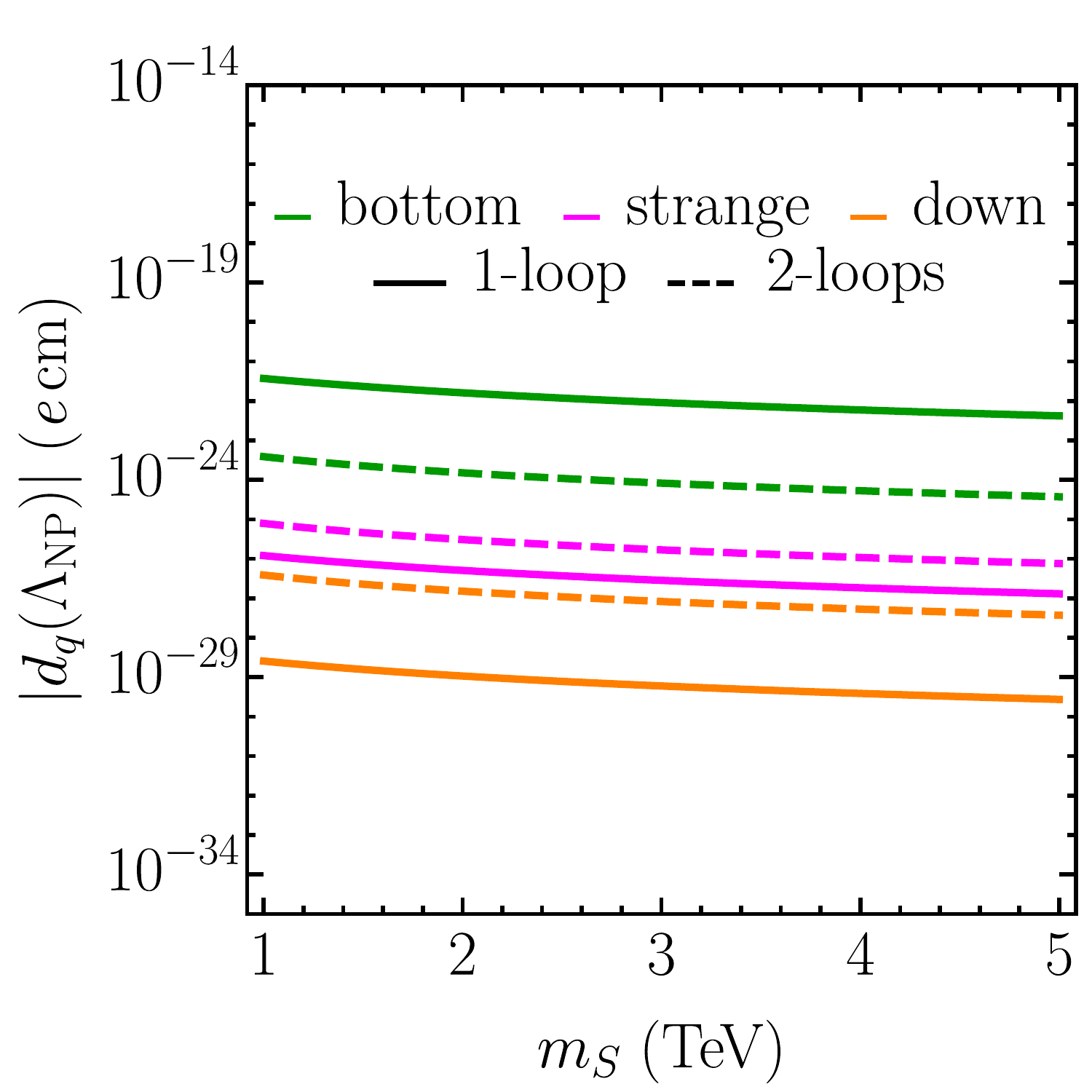} \includegraphics[width=0.46\textwidth]{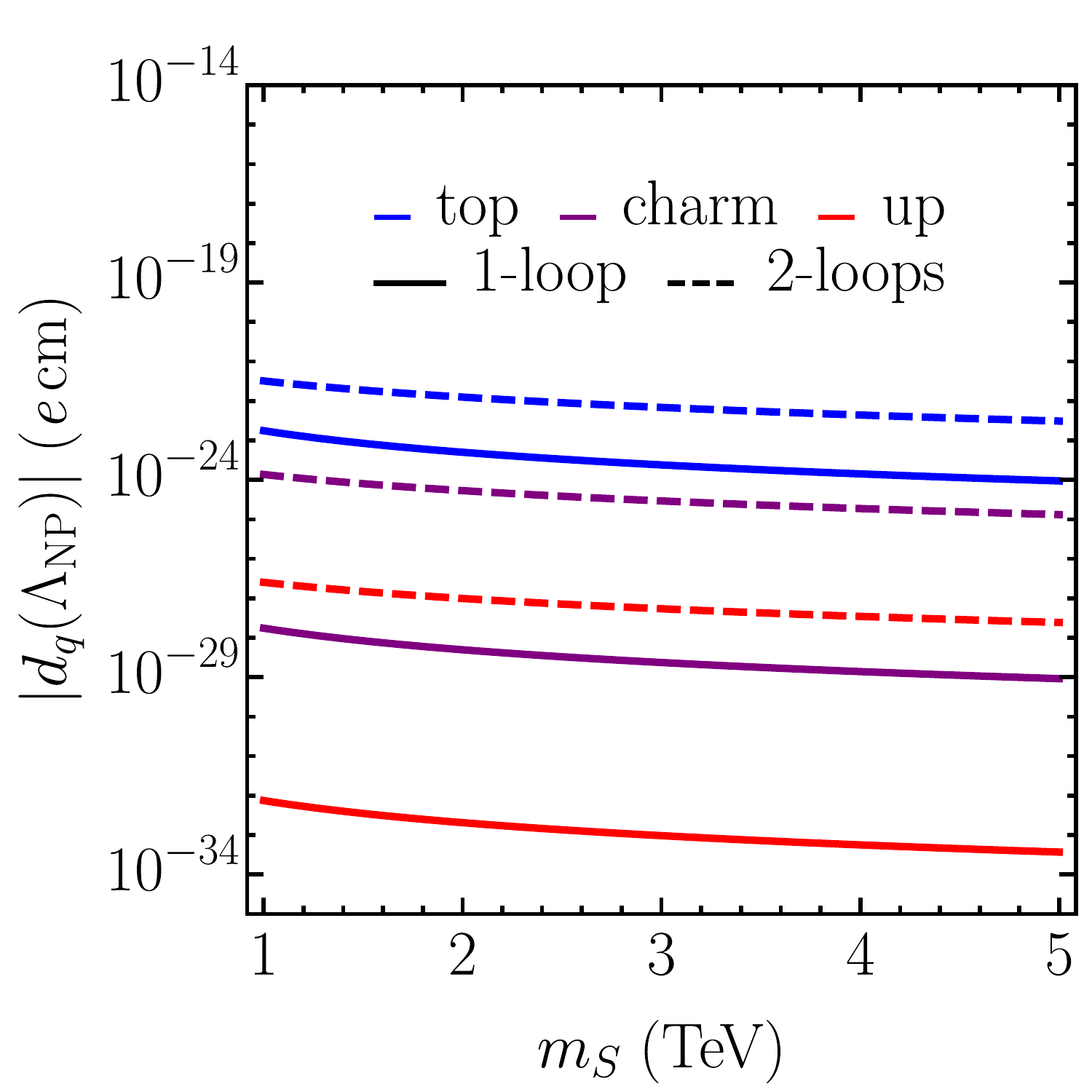}
    \\
    \includegraphics[scale=0.45]{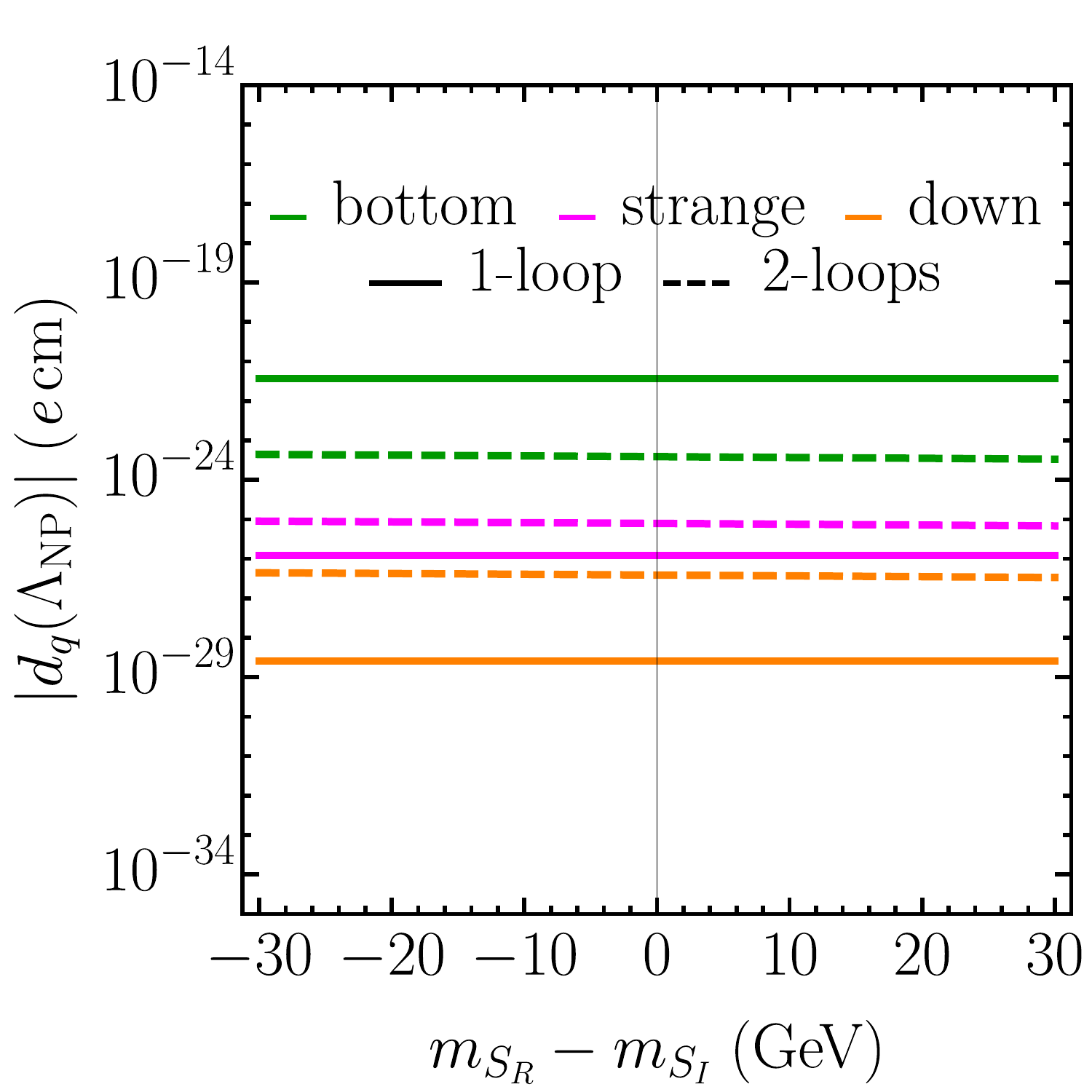}
    \includegraphics[scale=0.45]{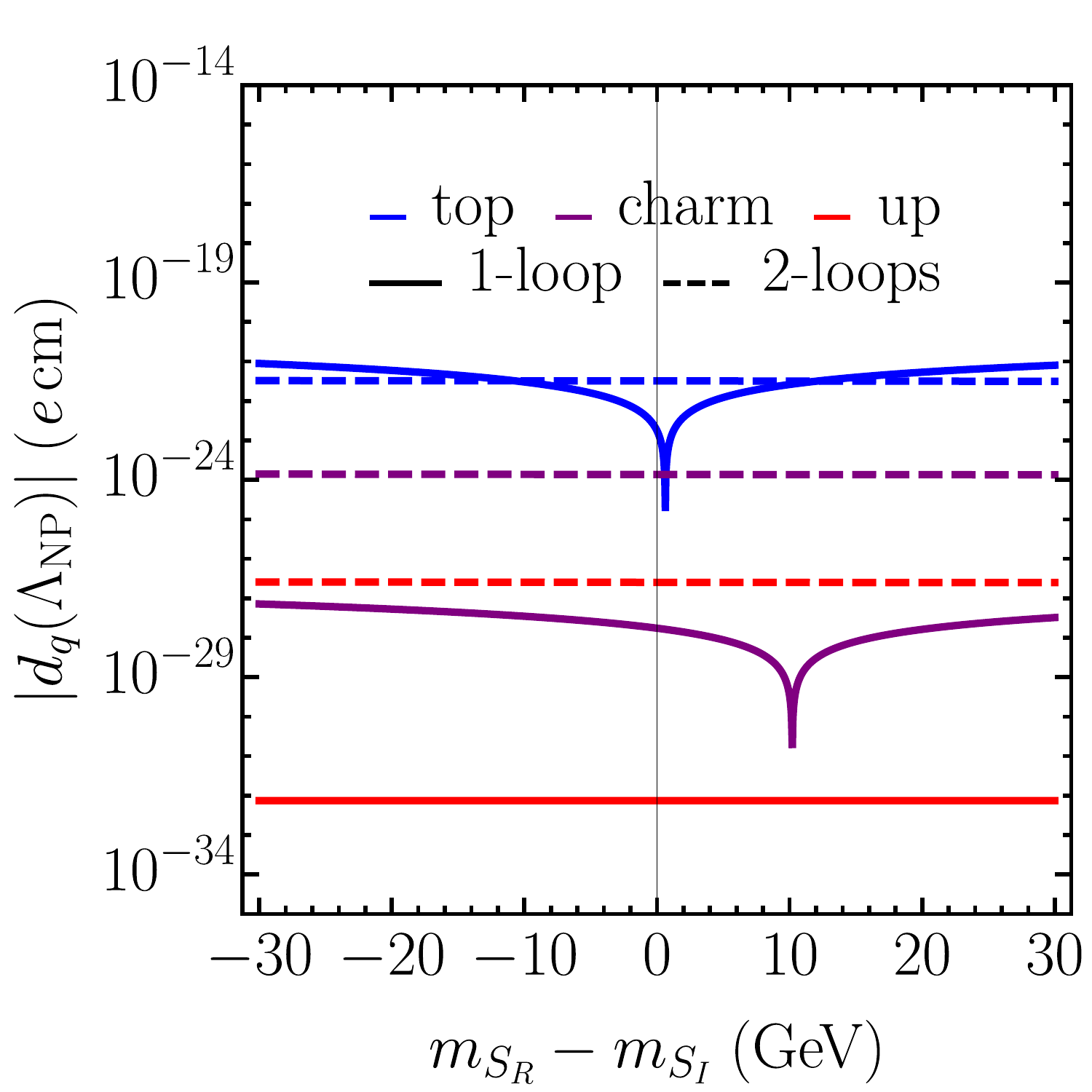}
    \caption{Comparison of the 1- and 2-loop contributions to the EDM of the up-type quarks (left) and down-type quarks (right), as a function of the mass of the scalars $m_S$ (top) and the mass splitting of the neutral scalars $m_{S_R} - m_{S_I}$ (below). The Yukawa couplings have been fixed to $|\eta_U| = |\eta_D|= 1$, $ \arg(\eta_U)=\pi/4$, and $ \arg(\eta_D)=0$. The masses have been fixed to $m_S^+ = m_{S_R} = m_{S_I} = m_S$, and $m_{S^+} =  m_{S_I} =1\,\text{TeV}$ in the top and bottom panels, respectively.
    }
    \label{fig:EDM_heavy_quarks}
\end{figure}

Before moving on with the phenomenological analysis to compare the model predictions against current experimental upper limits, it is worthwhile to compare the size of the Barr-Zee contributions, obtained above, to the one-loop contributions presented originally in Ref.~\cite{Martinez:2016fyd}.\footnote{Note, however, that our results do not agree with this reference for one of the loop functions. See the discussion above for more details.} 

Due to the strong suppression from powers of the light quark masses, the light quark (C)EDMs are dominated by two-loop Barr-Zee diagrams. Even though these contributions include additional coupling constants and loop suppression factors, the top-quark Yukawa coupling, proportional to $m_t$, makes the two-loop diagrams the dominant contribution for light quark (C)EDMs. With the same argument, one would expect heavy quark (C)EDMs to be dominated by one-loop diagrams, as they are not suppressed neither by light quark masses nor by loop suppression factors. This hierarchy of contributions is illustrated in Figure \ref{fig:EDM_heavy_quarks} (top-left), where we see that $d_b^{\rm 1-loop} > d_b^{\rm 2-loop}$, and vice-versa for the strange and down quark, $d_s^{\rm 1-loop} < d_s^{\rm 2-loop}$ and $d_d^{\rm 1-loop} < d_d^{\rm 2-loop}$.

This pattern is generally expected in models where the Yukawa couplings of new scalars are proportional to the quark mass. However, in the MW model, this hierarchy of contributions is apparently not respected for up-type quarks, as shown in Figure \ref{fig:EDM_heavy_quarks} (top-right), where also the 2-loop contribution dominates for the top quark EDM, $d_t^{\rm 1-loop} < d_t^{\rm 2-loop}$. This counter-intuitive behaviour can be explained by inspecting the expressions of $d_q$ at one-loop level. There, the CP-odd and CP-even neutral scalar contributions to the (C)EDMs have opposite signs, cancelling each other to a large extent, since the mass splitting $|m_{S_R^0} - m_{S_I^0} | \leq 30\,\text{GeV}$ as established by unitarity bounds \cite{Eberhardt:2021ebh}. In the limit where the masses are degenerate, $m_{S_R^0} = m_{S_I^0}$, only the charged scalar contribution is relevant at one-loop level, being $d_q^{\rm 1-loop}\propto m_q m_{q^\prime}^2 |V_{q q^\prime}|^2$. As a consequence, the top quark EDM $d_q^{\rm1-loop}$ is suppressed by the bottom quark mass, while only top-quark masses appear in $d_t^{\rm2-loop}$. The dependence on the mass splitting is explicitly shown in Figure \ref{fig:EDM_heavy_quarks} (bottom-right), where we see that as soon as $|m_{S_R^0} - m_{S_I^0} |$ deviates from zero, the neutral scalar contribution dominates and the expected hierarchy with $d_t^{\rm 1-loop} > d_t^{\rm 2-loop}$ is recovered. Nevertheless, within the allowed range of  $|m_{S_R^0} - m_{S_I^0} |$, note that the one- and two-loop level contributions to the top quark are of similar size, at least for some regions of the parameter space. In this Figure, the dip on $d_q^{\rm 1-loop}$ away from $|m_{S_R^0} - m_{S_I^0} | = 0$ is due to the cancellation of the neutral and charged scalar contributions.

This feature is not reproduced for down-type quarks, in Figure \ref{fig:EDM_heavy_quarks} (bottom-left), since the charged-scalar contribution dominates in all the range of masses. The reason for this is the enhancement from the heavier mass of the quark running in the loop. Namely, $d_b^{\rm 1-loop}\propto m_t^2$ and $d_s^{\rm 1-loop} \propto m_c^2$. As a consequence, at one-loop level, the bottom and strange quark EDMs are in fact larger than their up-type partners.

\section{Phenomenological analysis} \label{sec:pheno}

With all the relevant contributions of the coloured scalars to the EDM of hadrons obtained, we can study the constraints that these observables impose on the model. Currently, there is no direct limit on the EDM of the proton and it will not be used for this analysis. Furthermore, as commented previously, the implications from the neutron and mercury EDM on the MW model are extremely similar. Therefore, in order to provide clearer limits, we will only consider the direct limit on the neutron EDM in the numerical analysis. The values for the hadronic and nuclear matrix elements of Eq.~\eqref{eq:edm_fun} have been set to the central values. Of course, with only one observable and a total of seven parameters,
$$
|\eta_U|,\, |\eta_D|,\, \text{arg}(\eta_U),\,\text{arg}(\eta_D),\, m_{S_R},\, m_{S_I}, \text{ and } m_{S^\pm},
$$
we do not provide a complete phenomenological analysis here, but just a brief study showing the potential of EDM observables to constrain the parameter space of the model. To this end, the interplay of the model parameters appearing in the neutron EDM is discussed, comparing its limits to those imposed by other powerful observables studied in the literature. A global fit with all the relevant observables for the CP-violating MW model, although interesting, is beyond the scope of this paper. 

To reduce the total number of free parameters and provide some sensible plots, we will first assume that the mass of the scalars is degenerate, $m_{S_R}=m_{S_I}=m_{S^\pm}=m_S$. This assumption is completely reasonable given the constraints on the mass splitting found in Ref.~\cite{Eberhardt:2021ebh}. In addition, we will show here the results for $|\eta_U|=1$, which is below the maximum allowed value found in  Ref.~\cite{Eberhardt:2021ebh} for masses of the coloured scalars of a few TeVs. If $|\eta_U|=0$, the contributions with the top quark Yukawa coupling studied here vanish, and only the suppressed diagrams with bottom-quark propagators running in the loops contribute.

\newpage
\subsection{Neutron EDM predictions}

\begin{figure}[h!]
    \centering
    \includegraphics[scale=0.6]{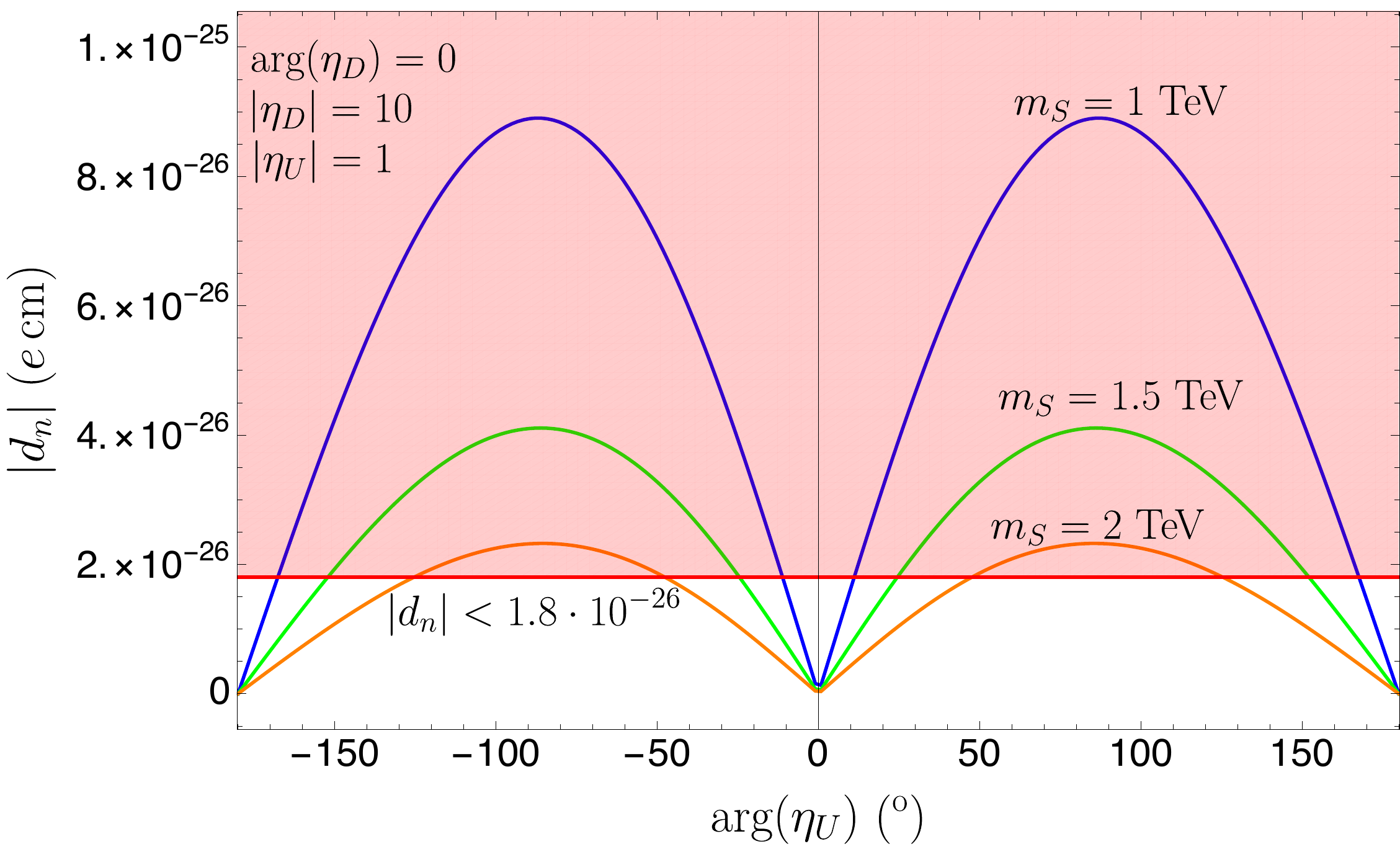}
    \caption{Electric dipole moment of the neutron as a function of the complex phase of $\eta_U$. The shaded region is excluded by the current experimental limit. }
    \label{fig:1D_EDM_alphaU}
\end{figure}

\begin{figure}[h!]
    \centering
    \includegraphics[scale=0.6]{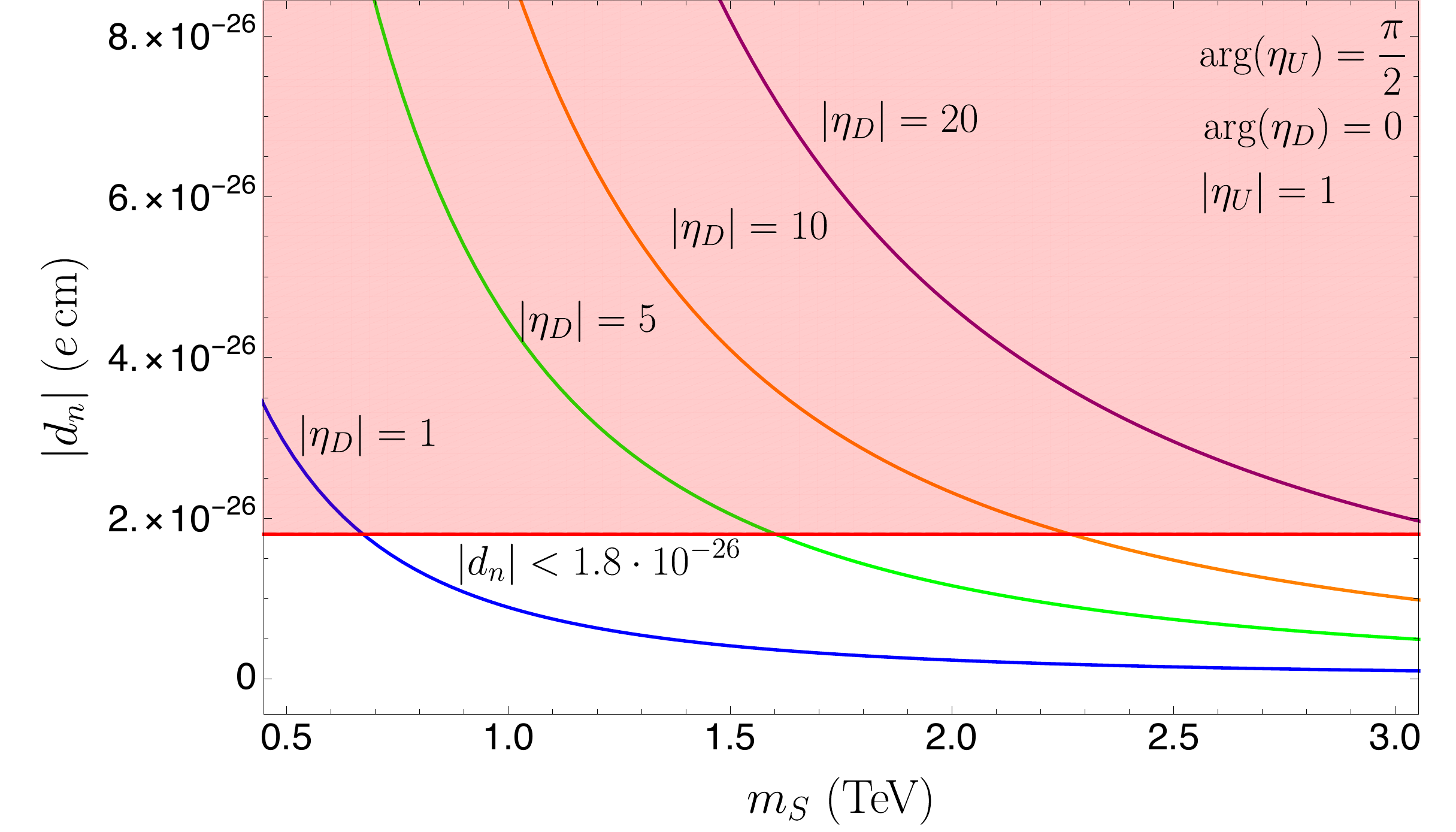}
    \caption{Electric dipole moment of the neutron as a function of the mass of the coloured scalars, $m_S$ (all scalar masses are fixed to the same value). The shaded region is excluded by the current experimental limit.}
    \label{fig:1D_EDM_mS}
\end{figure}

An intuitive way to see the effect of the neutron EDM bound on the MW model is to compare its prediction, as a function of the model parameters, with the current experimental limit. Besides giving clues on the allowed size of the model parameters (studied in more depth in Section~\ref{sec:phenoparams}), this also allows to evaluate the effect of future neutron EDM limits on this model.
In Figure~\ref{fig:1D_EDM_alphaU} we show the size of the neutron EDM as a function of $\text{arg}(\eta_U)$, fixing $|\eta_U|=1$, $|\eta_D|=10$ and $\text{arg}(\eta_D)=0$, for different values of the mass of the coloured scalars. Here we see how a strong constraint for $\text{arg}(\eta_U)$ can be obtained even for masses of the scalars around 1.5 TeV, using the current experimental limits for the neutron EDM. It is also interesting to look at the possible constraints on the scalar masses, obtained by fixing the other parameters. This can be seen in Figure~\ref{fig:1D_EDM_mS} where we have fixed $\text{arg}(\eta_U)=\pi/2$ (which gives the strongest contribution to the neutron EDM), $|\eta_U|=1$ and $\text{arg}(\eta_D)=0$, and we have varied $|\eta_D|$. Here we can see how for reasonable values of $|\eta_D|$, the mass of the scalars could be constrained to be higher than 3 TeV, far beyond the current experimental limit from direct searches.

\FloatBarrier

\clearpage

\subsection{Constraints on the model parameters } \label{sec:phenoparams}

\begin{figure}[h!]
    \centering
    \includegraphics[scale=0.81]{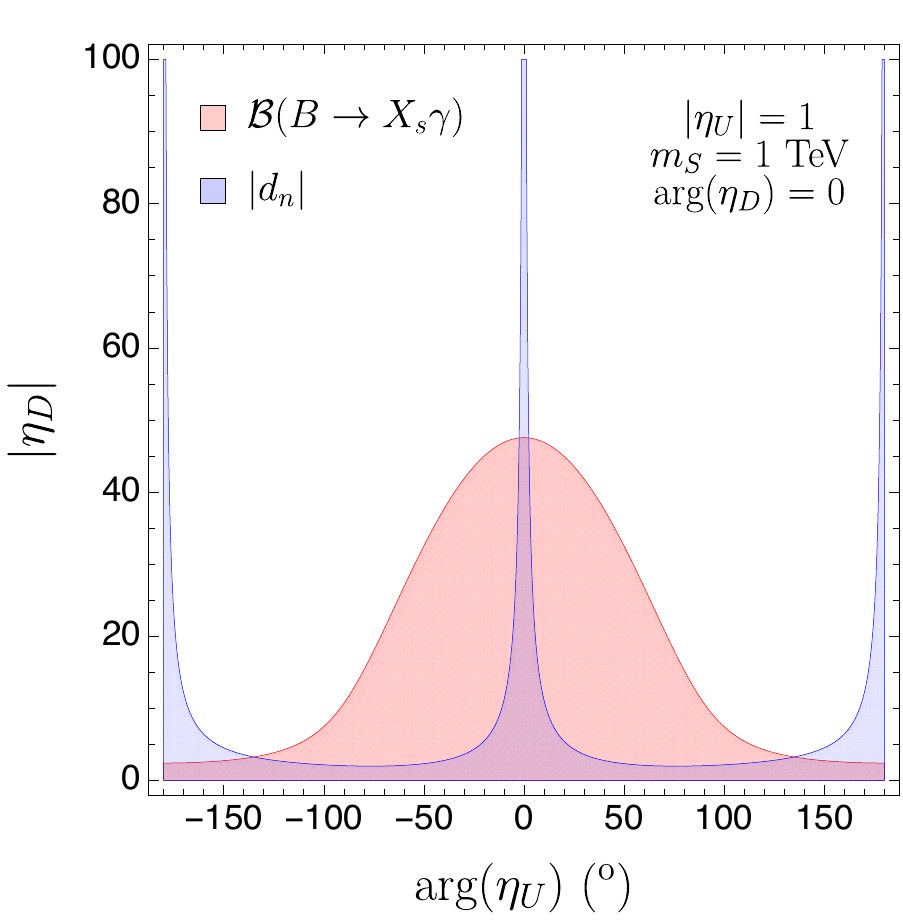}
    \includegraphics[scale=0.81]{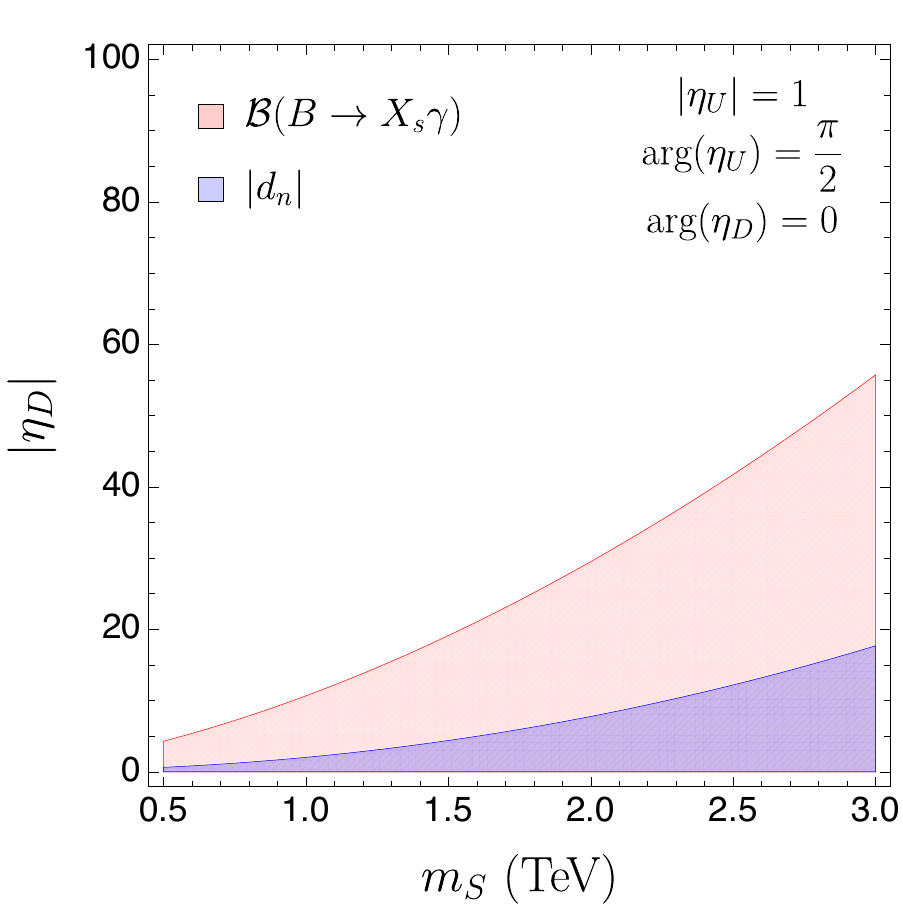}\\
    \includegraphics[scale=0.81]{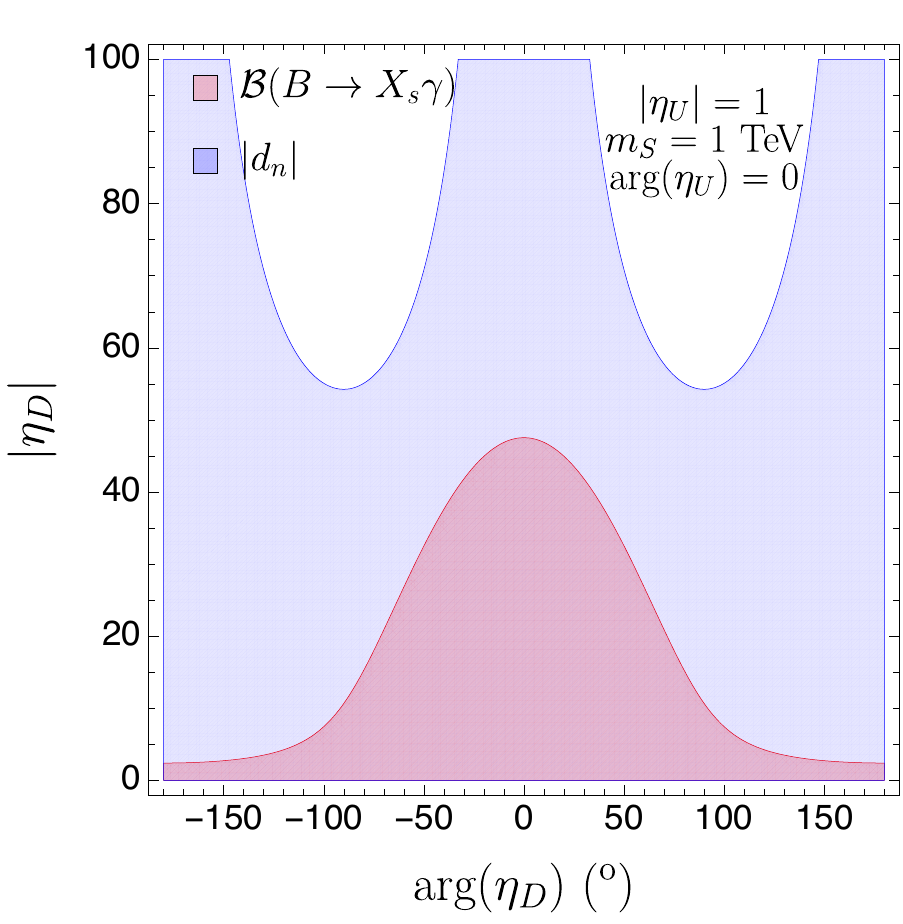}
    \includegraphics[scale=0.81]{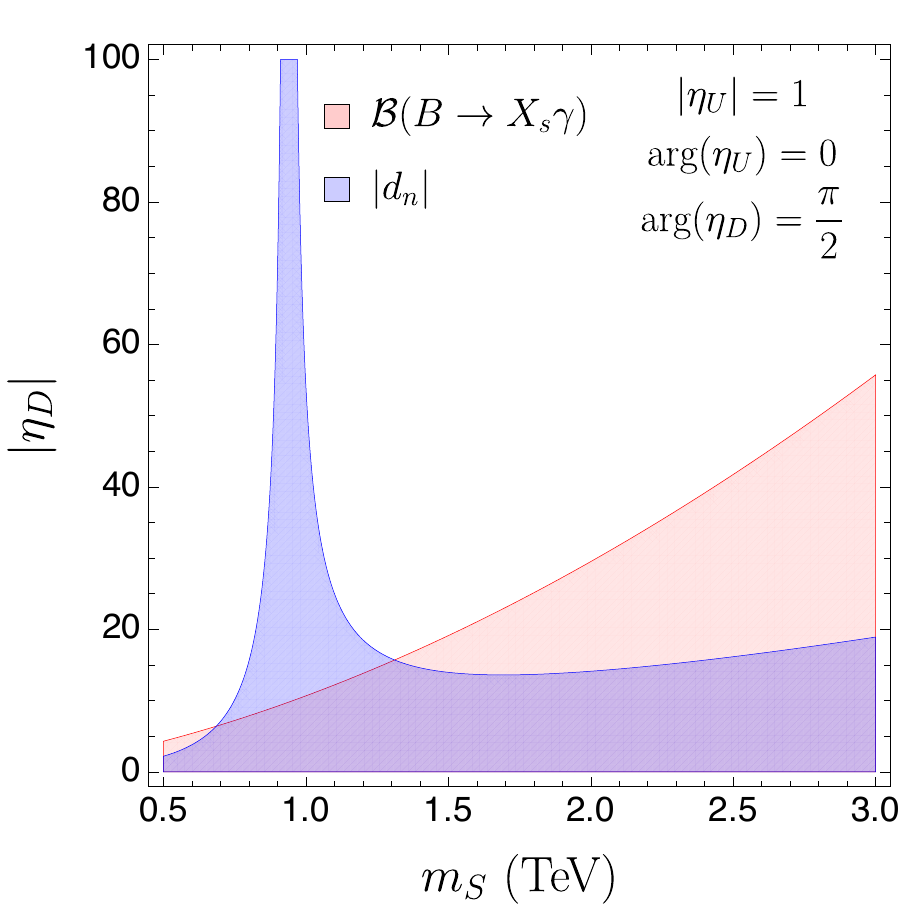}
    \caption{ Constraints on the parameter space of the MW model from the limits on the neutron EDM (blue) compared to those from $\mathcal{B}(B\rightarrow X_s\gamma)$ (red), calculated in Ref.~\cite{Cheng:2015lsa}. The coloured areas represent the allowed regions of parameter space. In the top panels one can see the large impact of the new EDM bounds on some regions of the parameter space. However, its restrictive power is diminished either when the absolute value of the Yukawa couplings is small enough, when the phases are $\arg(\eta_{U,D}) \approx0,\pm\pi$, or in the presence of cancellation effects as in $m_S \approx 1\,\text{TeV}$ (see text for details).
}
    \label{fig:2D_RegionPlot_EDM}
\end{figure}

To assess the restrictive power of the neutron EDM bounds, we can compare them to the most restrictive observables on the same planes of the parameter space. Following the global-fit analysis of Ref.~\cite{Cheng:2015lsa}, we chose the observable $\mathcal{B}(B\rightarrow X_s\gamma)$ as a benchmark to study the EDM restrictions.
This comparison is done in Fig.~\ref{fig:2D_RegionPlot_EDM}, where the region of the parameter space allowed by each observable is shown. In the following we describe these results, pointing out the main patterns in this figure.

We have fixed the phases $\text{arg}(\eta_U)=0$ ($\text{arg}(\eta_D)=0$) in the top (bottom) panels in order to study the effect of CP violation as coming from the down-type (up-type) Yukawa couplings. First, looking at the $|\eta_D|-\text{arg}(\eta_U)$ plane (top-left panel) one can see that the constraints from the neutron EDM are stronger than those of $\mathcal{B}(B\rightarrow X_s\gamma)$. The only exceptions lie in the vicinity of the values $\text{arg}(\eta_U) = 0,\pm\,\pi$, where $d_n$ vanishes and it cannot impose any restriction on the model parameters. Fortunately, an excellent experimental precision on $\mathcal{B}(B\rightarrow X_s\gamma)$, which is sensitive to both CP-violating and CP-conserving interactions, allows to restrict those directions even for these limiting values of the phases. This feature shows the power of combining the stringent experimental limits on EDMs with the complementary information from flavour observables.  Nonetheless, as in any interaction beyond the SM, when the absolute value of the new coupling is small enough, restrictions on other model parameters cannot be found. In our case and with the current experimental precision of the neutron EDM, this feature appears as an horizontal band at $|\eta_D|\lesssim1$, spanning along all the domain of $\text{arg}(\eta_U)$ in the top-left panel.

\FloatBarrier

\FloatBarrier

Regarding the $|\eta_D|-m_S$ plane, the constraints from $d_n$ are much stronger than those from $\mathcal{B}(B\rightarrow X_s\gamma)$ when $\text{arg}(\eta_U)=\pi/2$ and $\text{arg}(\eta_D)=0$ (top-right panel in Fig.~\ref{fig:2D_RegionPlot_EDM}). However, exchanging the values of the phases, $\text{arg}(\eta_U)=0$ and $\text{arg}(\eta_D)=\pi/2$ (bottom-right panel), an unconstrained direction appears for masses around $m_S\sim 0.9$ TeV (bottom-right panel). 
In this specific region of the parameter space, different contributions to the neutron EDM cancel out. In particular, the light quark (C)EDMs (through Barr-Zee diagrams) and the Weinberg operator (through the threshold contribution proportional to the bottom CEDM, $\tilde{d}_b$) have similar sizes. For $\text{arg}(\eta_U)=\pi/2$ and $\text{arg}(\eta_D)=0$, both contributions interfere constructively resulting in very stringent limits (top-right panel). Conversely, when the values of the phases are switched, $\text{arg}(\eta_U)=0$ and $\text{arg}(\eta_D)=\pi/2$, the Weinberg contribution flips sign, and the interference becomes destructive, preventing any constrain from the neutron EDM. 
To illustrate this dilution of the constraints, we fixed the mass value close to where the destructive interference is produced, $m_S= 1$ TeV, and plotted the allowed regions in the $|\eta_D|-\text{arg}(\eta_D)$ plane (bottom-left panel), keeping $\text{arg}(\eta_U)=0$. 
%

 

\section{Summary}
\label{sec:summary}

In this paper we have analysed the relevant contributions to the neutron EDM in the MW model. Expressions for the quark (C)EDM and Weinberg operators have been obtained, which can easily be generalised to other models with colour-octet scalars through the appropriate relations between the coupling constants.
In the case of the Weinberg operator, the neutral scalar contributions turn out to be irrelevant due to the cancellation between CP-odd and CP-even scalars, with the charged scalar contribution being completely dominant for this operator. In turn, only the neutral scalars produce sizable effects in the (C)EDM of light quarks through Barr-Zee type diagrams.

Using the current experimental limits on the neutron EDM, we found new stringent limits on the parameter space of the MW model when the Yukawa CP-violating phases are different from zero. Additionally, in the presence of strong cancellations between the contributions to the neutron EDM, or when the Yukawa phases are zero, we found a valuable complementarity of the neutron EDM with other flavour observables. In future works, the combination of these observables in a global-fit analysis will lead to the most stringent limits on the general CP-violating MW model.

\acknowledgments
 
We would like to thank Antonio Pich for useful discussions and the revision of this manuscript. We also thank Judith Plenter for her useful advice on the loop calculations and a careful revision of the manuscript. The work of HG is supported by the \textit{Bundesministerium f\"ur Bildung und Forschung} -- BMBF (Germany). The work of VM is supported by the FPU doctoral contract FPU16/0191 and the project FPA2017-84445-P, funded by the Spanish Ministry of Universities. JRV acknowledges support from MICINN and GVA (Spain).

\appendix

\section*{Appendices}

\section{Weinberg diagrams: straightforward calculation}
\label{app:wein}

In July 2021, the great physicist Steven Weinberg passed away, leaving behind a vast amount of contributions to physics and science. We owe to him 
the Electroweak theory, a keystone of modern particle physics, as well as many other outstanding contributions across many areas of theoretical physics. 
We see some of his footprints in our analysis of EDMs as well: he formulated the CP-odd three-gluon operator, in Eq.~\eqref{eq:opEFF}, and its two-loop leading contribution through the exchange of a scalar field. 
In his original paper~\cite{Weinberg:1989dx}, he provided the $h(r)$ loop function, while the full expression was obtained by D. Dicus in Ref.~\cite{Dicus:1989va}, who referred to the computation of this diagram as \textit{straightforward}.

At various points in the calculation we found, however, that equally justified choices of parameterisation or approximations can take the parametric integral off the \textit{straight} track towards this simple analytic expression. To our knowledge, these technical details are not found in the literature in a comprehensive summary. To facilitate the reproducibility of this analytical shape, we describe these details in the following. Following the same procedure, we arrived at the expression of $g(r)$, which is surprisingly simple as well, and at the cancellation of the diagram in Figure~\ref{fig:twoloopWeinbergNeut}~(c).

\begin{figure}[h]
	\centering
		\centerline{ \includegraphics[scale=1]{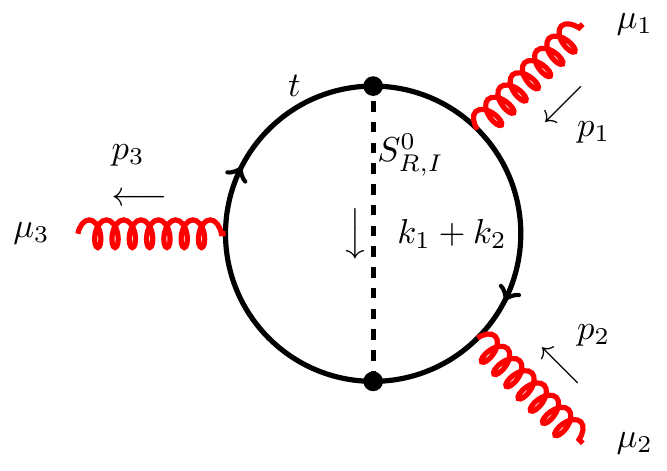} }    
	\caption{Diagram contributing to the Weinberg operator proportional to the loop function $h(r)$. The momentum directions are shown to facilitate the reproducibility of our results.}
	\label{fig:WeinbergMomentumTags}
\end{figure}

The Dirac trace of this two-loop amplitude contains up to eight $\gamma^\mu$, and two $\gamma_5$ matrices. Since the final result is finite, the traces with $\gamma_5$ can be solved with the usual relation ${\rm Tr}(\gamma^\mu \gamma^\nu \gamma^\rho \gamma^\sigma \gamma_5) = 4 i \varepsilon^{\mu \nu \rho \sigma}$. The $\mathcal{CP}$-violating parts of this amplitude are proportional only to the index structures with Levi-Civita tensors 
$$
\varepsilon^{\mu_1 \mu_2 \sigma \rho},~ \varepsilon^{\mu_2 \mu_3 \sigma \rho},~\varepsilon^{\mu_1 \mu_3 \sigma \rho} ,~\text{and}~\varepsilon^{\mu_1 \mu_2 \mu_3 \sigma},
$$
where the indices $\sigma$ and $\rho$ are contracted with external momenta.
To ease the calculation, it is convenient to select only one of these linearly-independent structures, as the final result is independent of this choice. The directions of the internal loop momenta were chosen as in Figure~\ref{fig:WeinbergMomentumTags}. With this, only three propagator denominators contain external momenta, which are small compared to the heavy mass $M$. Thus, we can expand these denominators in powers of $(p^2/M^2)$ with
\begin{align}
\frac{1}{(k_i+p)^2-M^2}=&
\frac{1}{k_i^2-M^2}
\left[
1-\frac{p^2 +2(p\cdot k_i) }{k_i^2-M^2}+\frac{4 (p \cdot k_i)^2 }{(k_i^2-M^2)^2}
\right] 
+{\cal O}(p^4/M^4)~,
\end{align}
and carefully removing higher-order terms after the expansions. Once the denominator is free of external momenta $p$, the tensor integrals with an odd number of open indices vanish, and, for the rest, we can apply the identity $k_i^\mu k_j^\nu \to (k_i \cdot k_j /D) g^{\mu\nu}$. The resulting master integrals have the shape
\begin{align}\label{eq:masterIntegralWeinberg}
\mathcal{W}_{\left\lbrace 00;~10;~11 \right\rbrace}(\alpha,\beta,\gamma;a,b,c)\equiv \int \frac{\text{d}^D k_1}{(2\pi)^D}\,\int \frac{\text{d}^D k_2}{(2\pi)^D}\,\frac{\left\lbrace 1;~k_1\cdot k_1;~k_1\cdot k_2\right\rbrace }{ \left(k_1^2 - a^2\right)^\alpha \left(k_2^2 - b^2\right)^\beta \left((k_1 \pm k_2)^2 - c^2\right)^\gamma  }~.
\end{align}
To re-express the $\mathcal{W}$ functions in terms of Feynman parameters, one must use Feynman parameterisation of two denominators at a time, and the standard Wick rotation to sequentially integrate over the loop momenta $k_1$ and $k_2$. The solution to the master integral that leads to the desired analytical shape of $h(r)$ reads 
\begin{align}
    \mathcal{W}_{\delta,\epsilon}(\alpha,\beta,\gamma;a,b,c)=\frac{(-1)^{D/2-\alpha-\gamma+\delta}}{(4\pi)^D}\frac{\Gamma(\alpha+\beta+\gamma-D-\delta)\Gamma(\delta+D/2)}{\Gamma(\alpha)\Gamma(\beta)\Gamma(\gamma)\Gamma(D/2)}w_{\delta\epsilon}(\alpha,\beta,\gamma;a,b,c),
\end{align}
where
\begin{align}
    w_{\delta\epsilon}(\alpha,\beta,\gamma;a,b,c)=&\int_0^1\,\text{d}x\,\int_0^1\,\text{d}y\;x^{D/2-\gamma-1}(1-x)^{\beta+\epsilon+1-D/2}y^{\alpha-1}(1-y)^{\beta+\gamma-D/2-1}\\
    \nonumber&\times \left(a^2y+(1-y)\frac{b^2x+(1-x)c^2}{x(1-x)}^{D-\alpha-\beta-\gamma+\delta}\right).
\end{align}

Selecting the tensor structure $\varepsilon^{\mu_1 \mu_2 \sigma \rho}$, only the function $\mathcal{W}_{11}$ contributes to this amplitude in the diagram of Figure~\ref{fig:WeinbergMomentumTags}. Finally, the three permutations of this diagram, obtained by rotating the internal scalar propagator by $120^o$ at a time, are obtained from the first diagram by renaming the indices and external momenta. This step is crucial to analytically cancel the divergencies of different parametric integrals. To obtain the Wilson coefficient, the fundamental amplitude must be matched onto the effective one, that reads~\cite{Dicus:1989va},
\begin{align} \label{eq:effectiveAmplitudeWeinberg}
i\mathcal{M}_{\rm eff} = &- \frac{2}{3}\, f_{abc}\, g_s\, w \,\varepsilon^{\mu_1} (p_1) \,\varepsilon^{\mu_2} (p_2)\, \varepsilon^{\mu_3} (-p_1 - p_2) \\ \nonumber
&\left[ (p_1 - p_2)_{\mu_3}\, \varepsilon_{\mu_1 \mu_2 \sigma \rho} + 2\, (p_{1~\mu_2}\, \varepsilon_{\mu_1 \mu_3 \sigma \rho} + p_{2~\mu_1}\, \varepsilon_{\mu_2 \mu_3 \sigma \rho})
\right] \,p_1^\sigma\, p_2^\rho~.
\end{align}
To develop these expressions, we used the help of the open-source packages \textsc{FeynArts} and \textsc{FeynCalc}~\cite{Shtabovenko:2020gxv,Kublbeck:1990xc,Shtabovenko:2016sxi}.

\section{Barr-Zee diagrams}
\label{app:BarrZee}

To simplify the calculation of the Barr-Zee diagrams, the two loops can be computed sequentially. The loop attached to the external photon (gluon) shall be obtained first. The result, in terms of Feynman integrals, can be written in the shape~\cite{Ilisie:2015tra}
\begin{equation}\label{eq:effvert}
i \Gamma^{\mu\nu}_{g,\gamma}=i (g^{\mu\nu} k\cdot q - k^\mu q^\nu) S_{g,\gamma} + i \epsilon^{\mu\nu\alpha\beta} k_\alpha q_\beta \widetilde{S}_{g,\gamma}~,
\end{equation}
where $q$ is the momentum of the external photon (gluon) and $k$ that of the off-shell gauge boson. The scalar functions $S$ and $\widetilde{S}$ encode all the relevant information of the different diagrams. The effective vertex of the dominant contributions to the (C)EDM, and the corresponding scalar form factors, read

\vspace{0.4cm}

\mpdiag{ \includegraphics[scale=1]{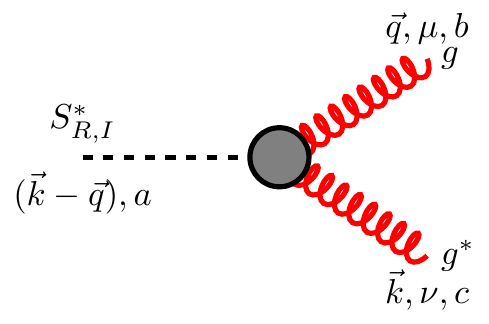}}{

\begin{align*}
S_g=&g_s^2\frac{m_t^2}{16\pi^2 v} d^{a b c} \text{Re}(\eta_U) \int_0^1 {\rm d} x \frac{2x^2-2x+1}{k^2(x-1)x+m_t^2}~,\\
\widetilde{S}_g=&-g_s^2\frac{m_t^2}{16\pi^2 v} d^{a b c}\text{Im}(\eta_U) \int_0^1 {\rm d} x \frac{1}{k^2(x-1)x+m_t^2}~,
\end{align*}

}

\mpdiag{ \includegraphics[scale=1]{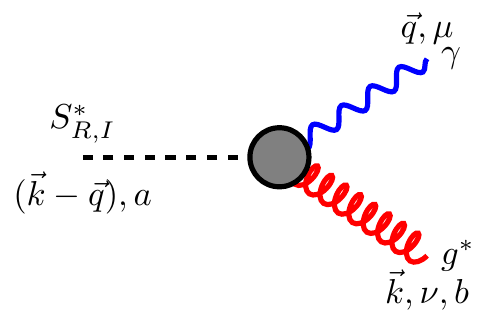}}{
	
\begin{align*}
S_\gamma=&e\,g_s\frac{m_t^2}{12\pi^2 v } \delta^{a b} \text{Re(}\eta_U\text{)}  \int_0^1 {\rm d} x \frac{  2 x^2- 2 x +1}{k^2 (x-1) x  + m_t^2}~,\\
\widetilde{S}_\gamma=&-e\,g_s\frac{m_t^2}{12\pi^2 v } \delta^{a b} \text{Im(}\eta_U\text{)}  \int_0^1 {\rm d} x\frac{  1}{ k^2 (x-1) x  + m_t^2}~.
\end{align*}

	}

\vspace{0.4cm}
	
\begin{figure}[t]
	\centering
	\subfloat[][]{\includegraphics[scale=1]{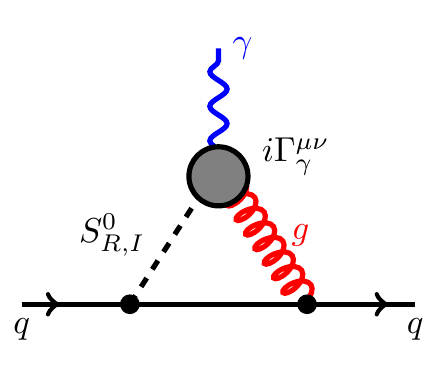}}
	\subfloat[][]{\includegraphics[scale=1]{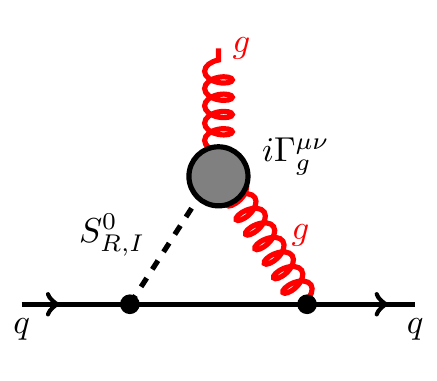}}
	\caption{Second loop of the Barr-Zee contributions to the quark EDM (a) and CEDM~(b).}
	\label{fig:secondloop}
\end{figure}

Only the top quark Yukawa coupling gives a sizeable contribution to this vertex. Thus, we only considered top quarks running in the \textit{inner} loop, as shown in Figure~\ref{fig:barr_zee}.
To arrive to this result, we use Feynman parametersiation in the shape of Eqs.~(6.23) and (6.25) of the detailed guide for computations~\cite{Ilisie:2016jta}. Furthermore, the photon (gluon) is assumed to be \textit{soft}, \textit{i.e.} $k\cdot p \to 0$, following the arguments of Ref.~\cite{Ilisie:2015tra}. 
Once the expressions for the first loop are parametrised as in Eq.~\eqref{eq:effvert}, this effective vertex is plugged in the second loop (Figure~\ref{fig:secondloop}), rewriting the denominator $k^2 (x-1)x + m_t^2$ as another propagator with momentum $k$. 
Then, the integrals over $k$ can be identified in terms of Passarino-Veltman functions. Expanding the result in powers of $(m_q/M)$, where $M$ is a heavy mass and $q={u,d}$, only the first term is numerically relevant. 
In this way, we obtained the loop functions $\mathcal{F}$ and $\widetilde{\mathcal{F}}$, in terms of the Feynman parameter $x$, which comes from the \textit{inner loop}. To match the fundamental amplitude to the effective (C)EDM operator, it is convenient to express the Levi-Civita tensor in terms of products of gamma matrices, through the Chisholm identity.

\FloatBarrier


\bibliographystyle{JHEP}
\bibliography{references}

\clearpage

\pagebreak

\end{document}